\documentstyle[aps,multicol,psfig,amsfonts,preprint]{revtex}
\newcommand{\ds}{\rule[-1.5mm]{0mm}{4.5mm}\displaystyle}
\renewcommand{\bar}[1]{\overline{#1}}
\begin{document}   
\newcommand{\z}{\zeta}
\newcommand{\bra}{\langle}
\newcommand{\ket}{\rangle}
\newcommand{\gap}{\hspace*{3pt}}
\newcommand{\bone}{\mbox{\em 1\hspace{-5.81truept}1\hspace{-5.81truept}1\hspace{-5.81truept}1\hspace{-5.81truept}1\hspace{-5.81truept}1}}
\parindent = 0pt
\bibliographystyle{prsty}
\draft   
\tightenlines
\widetext  
\title{ 
Statistical properties of eigenvectors  in
non-Hermitian Gaussian random matrix
ensembles
}   
 \author{B. Mehlig\footnote{Present address: Max-Planck-Institute 
                            for the physics of complex systems,
                            N\"othnitzer Str. 38, 01187 Dresden, Germany.}
         and J. T. Chalker}
 \address{\mbox{}Theoretical Physics, University of Oxford, 1 Keble Road, 
 OX1 3NP, United Kingdom}
\date{\today}
\maketitle{ } 
\begin{abstract}   
Statistical properties of eigenvectors in non-Hermitian
random matrix ensembles are discussed, 
with an emphasis on correlations between left and right eigenvectors.
Two approaches are described. One is an exact calculation 
for Ginibre's ensemble, in which each matrix element is an
independent, identically distributed Gaussian complex random variable.
The other is a simpler calculation using $N^{-1}$ as an expansion parameter, 
where $N$ is the rank of the random matrix: this is applied to Girko's ensemble. Consequences of eigenvector correlations which may be of physical importance
in applications are also discussed. It is shown that eigenvalues are much more sensitive to perturbations than in the corresponding Hermitian random matrix ensembles. It is also shown that, in problems with time-evolution governed by a non-Hermitian random matrix, transients are controlled by eigenvector correlations.
\end{abstract}   
\tableofcontents
\newpage

\section{Introduction}
Hermitian random matrices have been very successfully used
to model Hamiltonian operators of closed quantum systems \cite{meh67}.
In many cases, this has lead to a quantitative description
of features such as spectral fluctuations 
in classically chaotic quantum systems and in disordered
quantum systems in the metallic regime \cite{falk95}.
Within this approach it is also possible
to describe the statistical properties of
wave functions and matrix elements in such systems.
Random matrices
are also of great importance in many other areas of physics
in which they are not constrained to be Hermitian \cite{gin65,gir85}.
These include:
the dynamics
of neural networks \cite{som88},
the quantum mechanics of open
systems \cite{haa92}, classical diffusion
in random media \cite{cha97} and in population
biology \cite{nel98}, and modelling the statistical
properties of flux lines in superconductors
with columnar disorder\cite{hat96,efe97a,efe97b,gol98,hat98b}.
Recently, in connection with these
problems, spectral properties of non\--Her\-mi\-ti\-an
random matrices and operators have been studied
in great detail (see  for instance
\cite{gin65,gir85,cha97,nel98,fyo98,jan97,zee97,leh91,mud98}).

In the context of fluid dynamics
it is well known \cite{tref93,far96,tref97} that
systems governed by non-Hermitian
evolution operators exhibit striking
features. First, such systems are
particularly sensitive to perturbations.
Second, these systems can exhibit pseudo-resonances
at which the system reacts strongly
to an external perturbation although
the excitation frequency is not
close to any of the frequencies
of the internal modes.
Third, 
non-Hermiticity can give
rise to interesting
transient features in
time evolution.
Such features cannot be understood
solely in terms of the spectrum of the
evolution operator. While
the eigenvalues of the evolution operator
determine the long-time behavior, transients and
sensitivity to perturbations, in particular,
are determined by the properties of the
corresponding eigenvectors. 

In this paper 
we  quantify the
statistical properties of the
eigenvectors of random non-Hermitian matrices
and examine to what extent
enhanced sensitivity to perturbations and
transients in time evolution are present in
random systems described by  non-Hermitian
operators, such as  Fokker-Planck operators
\cite{cha97} or projected Hamiltonians
\cite{haa92}. 
We report
on results for two ensembles
of random matrices, namely
Ginibre's ensemble \cite{gin65}
and Girko's ensemble \cite{gir85}.
There are several reasons for
studying these ensembles. First,
characterizing the 
statistical properties of
eigenvectors in these cases is by itself
a problem of considerable interest: we
show that left and right eigenvectors
exhibit striking correlations, which depend strongly on
where in the spectrum the corresponding eigenvalues lie. 
Second, non-Hermitian operators
(such as the Fokker-Planck operator
governing classical diffusion
in a random velocity field \cite{cha97})
may be represented, in a finite system and
in an appropriate
basis, as random matrices. In general,
their matrix elements exhibit
certain structures and are much less
uniform than the matrices from the ensembles we investigate. 
Nevertheless, experience of universality in
random Hermitian problems gives reason to hope that
random matrix ensembles will provide a widely-applicable guide
to behavior.
Third, the high
symmetry of Ginibre's
ensemble (the matrix elements
are independently Gaussian distributed) allows for
an exact calculation which we present in detail. 
Separately, we develop an alternative, more general and simpler
approach to calculations, based on a perturbative evaluation of 
ensemble-averaged resolvents, using $N^{-1}$ as the expansion parameter,
where $N$ is the rank of the matrix. We apply this to Girko's ensemble, and
also assess its validity in Ginibre's ensemble 
by comparison with exact results.
Such approximate methods are particularly
important because they are easily extended
to the more general ensembles discussed
in \cite{cha97,nel98}.

The remainder of this article is organized as follows.
In section \ref{sec:not} we discuss
the formulation of the problem: we define
the ensembles of random non-Hermitian matrices
studied in subsequent sections, define the
densities of eigenvector overlaps that will
be the main objects of study in this paper,
and the corresponding Green functions.
This section establishes the notation
used subsequently.
In section \ref{sec:np}
we show how to derive exact results for
the statistical properties of
eigenvectors in Ginibre's ensemble
of non-Hermitian random matrices
of arbitrary matrix dimensions.
We also discuss simplifications which arise
in the large $N$ limit.
In section \ref{sec:p} we summarize
our approximate calculations, applying them to
Girko's ensemble. The eigenvector
correlators are calculated in
terms of two-point functions
which are obtained within
the self-consistent Born approximation.
This approach is appropriate in
the limit of large $N$,
and under certain additional
assumptions which are discussed
in this section. As a special
case, we check results found with this method for Ginibre's ensemble
against the exact results of section \ref{sec:np}.
The results obtained in sections
\ref{sec:ginibre} and \ref{sec:girko} are summarized and discussed in 
section \ref{sec:results}.
In section \ref{sec:imp} we
examine some consequences of eigenvector correlations that are likely to
be important in physical applications.
These are: the extreme sensitivity 
of eigenvalues to
perturbations; time
evolution governed by non-Hermitian random matrices;
and the nature of correlations between individual eigenvector
components in non-Hermitian random
matrix ensembles.
Finally, we summarize our conclusions in section \ref{sec:conc}.
An outline of some of these results has been published 
previously in a shorter communication \cite{cha98}.

\section{Formulation of the problem}
\label{sec:not}
\label{sec:problem}
\subsection{Ensembles of non-Hermitian matrices}
In the recent literature, a large number of
different ensembles of non-Hermitian random
matrices and operators have been discussed
\cite{som88,haa92,cha97,nel98,hat96,efe97a,%
efe97b,gol98,hat98b,fyo98,jan97,zee97,gin65,gir85}.
In the following, we restrict ourselves to
Ginibre's \cite{gin65} and Girko's ensembles \cite{gir85}
of non-Hermitian random matrices.

Ginibre introduced an ensemble
of random $N\times N$ matrices $J$ which have
complex elements $J_{kl}$ with
independently distributed real and imaginary parts $J_{kl}^\prime$
and $J_{kl}^{\prime\prime}$. The ensemble
is defined by the measure \cite{gin65}
(see also \cite{meh67})
\begin{equation}
\label{eq:ginibre}
P(J) \,{\rm d}J \propto \exp\Big (- \frac{1}{\sigma^{2}}\,
 \mbox{Tr}\, [J J^\dagger] \Big)  
\,\prod_{kl} {\rm d}J_{kl}^\prime {\rm d}J_{kl}^{\prime\prime}\,.
\end{equation}
Thus $\langle J_{kl} \rangle = 0$ and
$\langle J_{kl} \overline{J}_{kl} \rangle = \sigma^2$.
Here $\langle \cdots \rangle$ denote ensemble
averages and the overbar indicates
 complex conjugation.
The parameter $\sigma^2$ controls the density of
eigenvalues: in the limit $N\rightarrow\infty$,
the ensemble-averaged density (per unit area)
is $1/\pi\sigma^2$ within a disk in the complex plane, centered
on the origin and of radius $\sqrt{N}\sigma$.
Two different conventions are in use for the value of $\sigma^2$.
The choice $\sigma^2=1$ (as for instance in \cite{meh67,gin65})
results in a fixed density as $N\rightarrow\infty$.
Alternatively, the choice $\sigma^2=N^{-1}$ results in
a fixed support for the eigenvalue density
as $N\rightarrow\infty$.

Girko has considered the following generalization
of Ginibre's ensemble,
\begin{equation}
\label{eq:girko}
P(J)\,{\rm d}J \propto
\exp\left(-\frac{1}{\sigma^2}\frac{1}{1-\tau^2}
\mbox{Tr}\left[J J^\dagger -\tau\, \mbox{Re}JJ\right]\right)\,
\prod_{k,l} {\rm d}J_{kl}^\prime\,{\rm d}J_{kl}^{\prime\prime}\,,
\end{equation}
with $-1 \leq \tau \leq 1$. In this ensemble the non-zero cumulants are
\begin{eqnarray}
\label{eq:cumgirko}
\langle J_{kl}\,\bar{J}_{kl}\rangle &=& \sigma^{2}\,,
\hspace{1cm}\langle J_{kl}\,{J}_{lk}\rangle = \tau\sigma^{2}\,.
\end{eqnarray}
For $\tau = 0$, Ginibre's ensemble is recovered;
the case $\tau=1$ corresponds to Dyson's Gaussian
Unitary Ensemble \cite{meh67}, while
$\tau = -1$ describes an ensemble
of complex anti-Hermitian matrices.

\subsection{Densities of left and right eigenvectors}
\label{sec:denslrev}
The eigenvalues,
$\lambda_\alpha$, and left and right eigenvectors,
$\langle L_\alpha|$ and $|R_\alpha\rangle$, of the matrix $J$ satisfy
\begin{eqnarray}
\label{eq:ev}
J\,|R_\alpha\rangle   &=& \lambda_\alpha\,|R_\alpha\rangle\,,\\
\langle L_\alpha|\,J &=& \lambda_\alpha\,\langle L_\alpha|\,.
\nonumber
\end{eqnarray}
In general, the eigenvalues are complex numbers
$\lambda_\alpha = \lambda_\alpha'+{\rm i}\lambda_{\alpha}''$.
Except for a set of measure zero, they            are
non-degenerate. In this case the eigenvectors form
two complete, biorthogonal basis sets
with the normalization
\begin{equation}
\langle L_\alpha|R_\beta\rangle = \delta_{\alpha\beta}\,.
\label{eq:biorth}
\end{equation}
The closure relation is 
\begin{equation}
\sum_\alpha |L_\alpha\rangle\langle R_\alpha| = 1\,.
\end{equation}

We denote the Hermitian conjugates of $\langle L_\alpha|$
and $|R_\beta\rangle$ by $|L_{\alpha}\rangle$ and $\langle R_\beta|$,
so that, for example, $|L_{\alpha}\rangle$ satisfies 
$J^{\dagger} |L_{\alpha}\rangle = \lambda_\alpha^* |L_{\alpha}\rangle$.
Left  and right eigenvectors are generally not orthogonal
amongst themselves. On the contrary, scalar products can
vary significantly. This can have important physical
implications. For instance, it is well-known that non-orthogonality
of  eigenvectors can have an important bearing
of time evolution in systems governed
by non-normal operators \cite{tref97}. 

In the following we consider statistical properties of
scalar products of eigenvectors in ensembles of random
non-normal operators. 
We note that Eqs. (\ref{eq:ev}) and (\ref{eq:biorth})
allow for the following scale transformation
\begin{equation}
\label{eq:scaletr}
|R_\alpha\rangle \rightarrow c_\alpha\, |R_\alpha\rangle\,,\hspace{1cm}
\langle L_\alpha | \rightarrow \langle L_\alpha| \,c_\alpha^{-1}
\end{equation}
with arbitrary complex numbers $c_\alpha$:
we study only such combinations of eigenvectors
as are invariant under this scale transformation.
The simplest such combination of two eigenvectors
is trivial [see Eq. (\ref{eq:biorth})]. We hence
consider the combination
\begin{equation}
\label{eq:defO} 
O_{\alpha\beta} = \langle L_\alpha | L_\beta \rangle\,
                  \langle R_\beta  | R_\alpha\rangle\,.
\end{equation}
We calculate the mean value and discuss the distribution function
of this overlap matrix. 
Note that completeness implies the sum rule
\begin{equation}
\label{eq:sumrule}
\sum_{\alpha} O_{\alpha\beta} = 1\,. 
\end{equation}

It is convenient to
define local averages of diagonal and off-diagonal
elements of $O_{\alpha\beta}$,
\begin{eqnarray}
\label{eq:Odiag}
O(z) &=&  \Big\langle\sigma^2 \sum_\alpha O_{\alpha\alpha}
\,\delta(z-\lambda_\alpha)\Big\rangle\,,\\
\label{eq:Ooff}
O(z_1,z_2) &=&\Big\langle \sigma^2\sum_{\alpha\neq\beta} O_{\alpha\beta}
\,\delta(z_1-\lambda_\alpha) \,\delta(z_2-\lambda_\beta)
\Big\rangle\,.
\end{eqnarray}
Here, $z=x+{\rm i}y$ is a complex number with real and
imaginary parts $x$ and $y$
and $\delta(z)$ denotes a delta-function in both coordinates.
Correspondingly,
the density of states and the two
point function are
defined as 
\begin{eqnarray}
\label{eq:dnewp}
d(z) &=& \Big\langle \sigma^2 \sum_\alpha \delta(z-\lambda_\alpha)\Big\rangle\\
\label{eq:R2newp}
R_2(z_1,z_2) &=& \Big\langle \sigma^2 
\sum_{\alpha\neq\beta} \,\delta(z_1-\lambda_\alpha)\,
\delta(z_2-\lambda_\beta)\Big\rangle .
\end{eqnarray}

In order to characterize the overlap matrix
using Green functions, it is convenient
to introduce the density
\begin{equation}
\label{eq:defD}
D(z_1,z_2)  = \Big \langle \sigma^2 \sum_{\alpha,\beta} 
O_{\alpha\beta} \,\delta(z_1-\lambda_\alpha)\,\delta(z_2-\lambda_\beta)
\Big\rangle\,,
\end{equation}
which can be expressed in terms of $O(z_1)$ and $O(z_1,z_2)$
as 
\begin{eqnarray}
\label{eq:DOO}
D(z_1,z_2) &=&O(z_1)\,\delta(z_1-z_2) + O(z_1,z_2)\,.
\end{eqnarray}
Thus, information on the diagonal overlap
matrix elements may be extracted from 
the singular part of $D(z_1,z_2)$. The smooth
part conveys information
on the off-diagonal overlap matrix elements.

Finally we note that the sum rule (\ref{eq:sumrule})
implies the constraint for the density $D(z_1,z_2)$
\begin{equation}
\label{eq:sumruleD}
\int {\rm d}^2 z_2\, D(z_1,z_2) = d(z_1)
\end{equation}
where $d(z_1)$ is the density of states [Eq. (\ref{eq:dnewp})].

\subsection{Green functions and spectral densities}
We shall make use of the fact that 
the densities $d(z)$ 
and $D(z_1,z_2)$ may be expressed in
terms of ensemble averages of resolvents $(z-J)^{-1}$
and products of resolvents
$(z_1-J)^{-1}\,(\overline{z}_2-J^\dagger)^{-1}$.

The density of states $d(z)$, for example, by means of the relation
\begin{equation}
\label{eq:trick}
\delta(z) =  \frac{1}{\pi} \frac{\partial}{\partial \overline{z}}\,
\,\frac{1}{z}
\end{equation}
may be expressed as \cite{cha97,jan97}
\begin{equation}
\label{eq:dos}
d(z) = 
\frac{\sigma^2}{\pi}\frac{\partial}{\partial \bar z}\,  
\left\langle
\mbox{Tr}\,[ (z-J)^{-1}] \right\rangle\,.
\end{equation}
In Eqs. (\ref{eq:trick}) and (\ref{eq:dos})
\begin{equation}
\frac{\partial}{\partial \overline{z}}
= \frac{1}{2}\left(
\frac{\partial}{\partial x} + {\rm i}
\frac{\partial}{\partial y}\right)\,.
\end{equation}
Eq. (\ref{eq:trick}) replaces  the relation
$ \delta(E) = {\pi}^{-1} \mbox{Im}\,(E-i0^+)^{-1}$
which is applicable in problems for which  the Green functions
are analytic in the upper and lower complex half-planes.
Here, this is not the case. 

Similarly, the density $D(z_1,z_2)$ can be obtained 
 from a two-point function
\begin{equation}
\label{eq:densD}
D(z_1,z_2) = \frac{\sigma^2}{\pi^2} 
\frac{\partial}{\partial \bar z_1}\,  
\frac{\partial}{\partial  z_2}\,  
\left\langle  \mbox{Tr}\left [
(z_1-J)^{-1} \, (\bar z_2-J^\dagger)^{-1}\right] \right\rangle\,.
\end{equation}
This is most easily seen from
the spectral representation of the resolvent:
$(z-J)^{-1} = \sum_\alpha |R_\alpha\rangle\,
(z-\lambda_\alpha)^{-1}\,
\langle L_\alpha|$. We show in section \ref{sec:girko} how the
averages in Eqs. (\ref{eq:dos}) and (\ref{eq:densD})
can be calculated perturbatively,
using an expansion in powers of $N^{-1}$.

\section{Ginibre's ensemble}
\label{sec:np}
\label{sec:ginibre}
As pointed out in the introduction, Ginibre's ensemble
is a special case of Girko's family of ensembles of non-Hermitian
matrices. It is obtained by setting $\tau = 0$
in Eq. (\ref{eq:girko}) and is thus the 
ensemble of complex matrices with independent,
Gaussian distributed elements. In this special
case we are able to provide an exact calculation of 
the eigenvector correlators introduced
in section \ref{sec:problem}.

\subsection{Density-of-states and eigenvalue correlations}
\label{sec:dosginibre}

Eigenvalue correlations for the ensemble (\ref{eq:ginibre})
were first studied by Ginibre \cite{gin65}. The joint
probability distribution of the eigenvalues is
\begin{equation}
\label{eq:pz}
\label{eq:ginibrenewp}
P_N(\lambda_1,\ldots,\lambda_N)  = C_N
      \prod_{\mu < \nu} |\lambda_\mu-\lambda_\nu|^2\,  
      \exp\Big (-\frac{1}{\sigma^{2}}\,\sum_\mu |\lambda_\mu|^2\Big)
\end{equation}
with normalization 
$C_N = (N!\,\prod_{j=0}^{N-1} \pi\,j!\,\sigma^{2j+2})^{-1}$.
The eigenvalue density and the two-point function,
$d(z)$ and $R_2(z_1,z_2)$ [Eqs. (\ref{eq:dnewp}),(\ref{eq:R2newp})], 
may be calculated by  averaging  
$\delta(z-\lambda_1)$
and $(N\!-\!1)\, \delta(z_1-\lambda_{1})\,\delta(z_2-\lambda_2)$
with the weight $P_N$.
In the following we demonstrate briefly a 
way of performing the corresponding $\lambda$-integrals
which can be readily generalized to deal
with the integrals that arise in the calculation
of the eigenvector correlations 
[Sec. \ref{sec:lambda_int2}]. 
Making use of the fact that
\begin{eqnarray}
P_N(\lambda_1,\ldots,\lambda_N) 
&=& (\pi\, N! \,\sigma^{2N})^{-1}\,
\exp\Big(\!\!-\frac{\gap|\lambda_1|^2}{\gap\sigma^2}\Big)\,
 \prod_{m=2}^{N} 
   |\lambda_1-\lambda_m|^2 P_{N-1}(\lambda_2,\ldots,\lambda_{N})
\end{eqnarray}
we have  
\begin{eqnarray}
d(z) &=& N\sigma^2\,(\pi\, N!\,\sigma^{2N})^{-1}\, 
\exp\Big(\!\!-\frac{\gap|z|^2}{\gap\sigma^2}\Big)\,
      \int\!{\rm d}^2\lambda_2\cdots 
       {\rm d}^2\lambda_{N} \,P_{N-1}(\lambda_2,\ldots,\lambda_{N})
       \prod_{m=2}^{N} |z-\lambda_m|^2\,.
\end{eqnarray}
This can be written as
\begin{equation}
\label{eq:detdos}
d(z) = N\sigma^2\,
(\pi\,N!\,\sigma^{2})^{-1}\,
 \exp\Big(\!\!-\frac{\gap|z|^2}{\gap\sigma^2}\Big)\,
\det\left [
\begin{array}{ccccr}
d_{00} & d_{01} &  & &   0                      \\
d_{10} &        &\ddots  & &                          \\
 &\hspace*{-3mm}\ddots        &  & &                          \\
       &  &  &            & d_{N-3N-2}    \\
0      &        &  & d_{N-2N-3} & d_{N-2N-2}   
\end{array}
\right]
\end{equation}
with $d_{ij} = (\pi\, j!\,\sigma^{2j+4})^{-1} \int\!d^2\lambda\, 
\overline{\lambda}^i{\rule[0mm]{0mm}{4mm}\lambda}^j \,
|z-\lambda|^2\,\exp(-\sigma^{-2}\,|\lambda|^2)$.
Denoting the $(N\!-\!1\!)\times (\!N\!-\!1)$ determinant
in Eq. (\ref{eq:detdos}) by $D_{N-1}$ we  
derive
the recursion relation
\begin{equation}
D_{k+1} = (\sigma^{-2}\,|z|^2+k+1) D_k - \sigma^{-2}\,|z|^2\, k\, D_{k-1}.
\end{equation}
Using $D_1 = 1+\sigma^{-2}\,|z|^2$ and 
$D_2 = 2+2\,\sigma^{-2}\,|z|^2 + \sigma^{-4}\,|z|^4$,
we thus obtain
\begin{equation}
d(z) = \pi^{-1}\, \exp\Big(\!\!-\frac{\gap|z|^2}{\gap\sigma^2}\Big)\, 
\sum_{l=0}^{N-1}
\frac{\gap|z|^{2l}}{l!\,\sigma^{2l}}
\end{equation}
which corresponds to Eq. (51.1.32) in \cite{meh67}.
In the limit of large $N$,
with $\sigma^2 = N^{-1}$,
the density of states is 
\begin{equation}
\label{eq:doslargeN}
d(z) = \left\{
\begin{array}{ll}
\pi^{-1} & \mbox{for $|z| < 1$}\,,\\[0.2cm]
0        & \mbox{otherwise}\,.
\end{array}
\right .
\end{equation}

Similarly, we obtain for the two-point function 
\begin{eqnarray}
\label{eq:detR2}
R_2(z_1,z_2) &= & N\sigma^2\,
(\pi^2\, N!\, \sigma^6)^{-1}\,
|z_1-z_2|^2\,
\exp(-\sigma^{-2}\,|z_1|^2-\sigma^{-2}|z_2|^2)\,\\[0.5cm]
&\times&
\det\left [
\begin{array}{cccccr}
f_{00} & f_{01} &f_{02} &          &            &   0        \\
f_{10} & f_{11} &       & \ddots   &            &            \\
f_{20} & \phantom{\ddots}       &       &          &            &            \\
       & \ddots &       &          &            & f_{N-5N-3} \\
       &        &       &          & f_{N-4N-4} & f_{N-4N-3} \\
0      &        &       &f_{N-3N-5}& f_{N-3N-4} & f_{N-3N-3}   
\end{array}
\right]
\nonumber
\end{eqnarray}
with 
\begin{equation}
f_{ij} 
= \big(\pi \,(j\!+\!1)!\,\sigma^{2j+6}\big)^{-1} \int\!d^2\lambda\, 
\overline{\lambda}^{i}{\rule[0mm]{0mm}{4mm}\lambda}^{j} \,
|z_1-\lambda|^2\,|z_2-\lambda|^2
\exp\big(-\frac{|\lambda|^2}{\sigma^2}\big)\,.
\end{equation}
As before, we derive a recursion relation
for the $(N\!-\!2)\times (N\!-\!2)$ determinant
in Eq. (\ref{eq:detR2}).
This recursion relation simplifies considerably
when $z_2=0$. 
Denoting the determinant in Eq. (\ref{eq:detR2}) by $F_{N-2}$,
we have with $z_1=z$
\begin{equation}
F_{k+1} 
= (\sigma^{-2}\,|z|^2+k+2) F_k - \sigma^{-2}\,|z|^2\, (k+1)\, F_{k-1}\,.
\end{equation}
In this way we obtain, with $F_1 = 2+\sigma^{-2}\,|z|^2$
and $F_2 = 6+3\, \sigma^{-2}\, |z|^2 +\sigma^{-4}\, |z|^4$
\begin{equation}
R_2(z,0) = 
-\frac{1}{\pi^2\sigma^2}\, 
\exp\Big(\!\!-\frac{\gap|z|^2}{\gap\sigma^2}\Big)\,
\Big (1-\sum_{l=0}^{N-1} \frac{\gap|z|^{2l}}{l!\,\sigma^{2l}}
                                \Big  )
\end{equation}
which (for $\sigma^2 = 1$) is equivalent to (15.1.30) in \cite{meh67}
with $n=2$ and $z_2=0$. 
Moreover, in the limit of large $N$, 
with $\sigma^2 = N^{-1}$,
one finds that for $z_1 \neq z_2$ and $|z_1|,|z_2| < 1$,
the two-point function is constant \cite{meh67}
\begin{equation}
\label{eq:R2largeN}
N^{-1}R_2(z_1,z_2)
= \left \{
\begin{array}{ll}
\pi^{-2}
& \mbox{for $z_1 \neq z_2$ and $|z_1|,|z_2| < 1$},\\[0.2cm]
0        & \mbox{otherwise.}
\end{array}
\right .
\end{equation}

\subsection{Eigenvector correlations}
In this section we show how to obtain expressions
for correlations of eigenvectors in Ginibre's ensemble.
We start from (\ref{eq:Odiag}) and (\ref{eq:Ooff}),
perform calculations for general $\sigma^2$
but set $\sigma^2=N^{-1}$ in the final results
[see Eqs. (\ref{eq:resO11ca}) and (\ref{eq:resO12ca}) 
 below].

\subsubsection{Change of basis}
\label{sec:newbasis}
Since the fluctuations of
the eigenvectors and those
of the eigenvalues 
are correlated,
it is convenient to parameterize the matrix $J$
following Ref\,\cite{meha}, using a unitary transformation $U$
to bring it into upper triangular form,
\begin{equation}
{T} = {U}^\dagger {J} {U}
=
\left(
\begin{array}{llll}
\lambda_1 & T_{12}    & \cdots & T_{1N}\\
0         & \lambda_2 & \cdots & T_{2N}\\
\vdots    &           &        &\vdots\\
0         &           &        & \lambda_N
\end{array}
\right)\,.
\end{equation}
The ensemble requires $2 N^2$ coordinates.
Of these,  $2N$ are given by real and imaginary parts
of the eigenvalues $\lambda_k$, 
and $N(N-1)$ by real and imaginary parts
of the matrix elements $T_{kl}$.
The remaining $N(N-1)$ parameters $H_{kl}$ are 
as described by Mehta 
\cite{meha}.
The Jacobian of this transformation is proportional
to $\prod_{k<l} |\lambda_k - \lambda_l|^2$ and
thus depends on $\lambda_1,\ldots,\lambda_N$ only.
Note also that the eigenvector correlator $O_{\alpha\beta}$ is
invariant under the unitary transformation $U$. In this section,
$\langle L_\mu | $ and $|R_\nu\rangle$ will denote
left and right eigenvectors in the new basis.
Thus, $T|R_1\rangle = \lambda_1|R_1\rangle$
and $|R_1\rangle = (1,0,\ldots,0)^T$.
In keeping with Eq. (\ref{eq:biorth}), let
$\langle L_1| = (1,b_2,\ldots,b_N)$.
The coefficients $b_l$ can be determined by recursion:
From $\langle L_1| T = \lambda_1 T$ 
one has,  with $b_1=1$ and for $p>1$,
\begin{equation}
\label{eq:recursion1}
b_p = \frac{1}{\lambda_1-\lambda_p} \sum_{q=1}^{p-1} b_q
T_{qp}\,.
\end{equation}
The solution of this recursion relation is
\begin{eqnarray}
b_1 & = & 1\nonumber\\
b_2 & = & \frac{T_{12}}{\lambda_1-\lambda_2}\nonumber\\
b_3 & = &
      \label{eq:recsol1}
        \frac{T_{13}}{\lambda_1-\lambda_3} 
      + \frac{T_{12} T_{23}}{(\lambda_1-\lambda_2)(\lambda_1-\lambda_3)} \\
b_4 & = & 
        \frac{T_{14}}{\lambda_1-\lambda_4} 
      + \frac{T_{13} T_{34}}{(\lambda_1-\lambda_3)(\lambda_1-\lambda_4)}
      + \frac{T_{12} T_{24}}{(\lambda_1-\lambda_2)(\lambda_1-\lambda_4)}
            \nonumber\\
      &+&\frac{T_{12} T_{23} T_{34}}
      {(\lambda_1-\lambda_2)(\lambda_1-\lambda_3)(\lambda_1-\lambda_4)}
\nonumber\\
&\vdots &\nonumber
\end{eqnarray}

Eq. (\ref{eq:recsol1}) provides an explicit expression for
the correlator
\begin{equation}
\label{eq:O11}
O_{11} = \sum_{l=1}^N |b_l|^2
\end{equation}
in terms of the eigenvalues $\lambda_k$ and the matrix
elements $T_{kl}$ for $k < l$.

To calculate off-diagonal correlators one needs, in addition,
the eigenvectors $\langle L_2 |$ and $|R_2\rangle$.
Let $|R_2\rangle =  (c,1,0,\ldots,0)^T$
and, in keeping with Eq. (\ref{eq:biorth}),
$\langle L_2| = (0,1,d_3,\ldots,d_N)$.
Eq. (\ref{eq:biorth}) implies that $c=-b_2$.
Then $\langle L_2 | T = \lambda_2 T$ gives, with $d_1=0$ and
$d_2 = 1$,
\begin{equation}
\label{eq:recursion2}
d_p = \frac{1}{\lambda_2-\lambda_p} \sum_{q=1}^{p-1} d_q T_{qp}\,.
\end{equation}
This recursion relation is solved in the same way as
(\ref{eq:recursion1}) and
\begin{equation}
\label{eq:O12}
O_{12} = -\overline{b}_2\sum_{l=1}^N \overline{d}_l b_l
\end{equation}
provides a corresponding expression for $O_{12}$
in terms of the eigenvalues $\lambda_l$
and the matrix elements $T_{kl}$ ($k < l)$.

\subsubsection{Integration on $T_{kl}$}
It was shown in the previous section
how the correlators $O_{11}$ and $O_{12}$ may be expressed
in terms of the eigenvalues $\lambda_k$ and
the matrix elements $T_{kl}$ ($k < l$).
The Jacobian 
depends only on $\lambda_1,\ldots,\lambda_N$.
In calculating averages of the
the type (\ref{eq:Odiag}) 
and (\ref{eq:Ooff}),
the $N(N-1)$ parameters $H_{kl}$
mentioned in section \ref{sec:newbasis} can
thus be integrated out  and only
the integrals over $T_{kl}$ for $k< l$
and over $\lambda_k$ for $k=1,\ldots,N$
remain. These have the form
\begin{equation}
\int\!\prod_k {\rm d}^2\lambda_k\,\prod_{k<l} |\lambda_k-\lambda_l|^2\,
\prod_{k< l} {\rm d}^2 T_{kl}\,
\cdots \,\exp\left(-\frac{1}{\sigma^2}\sum_k |\lambda_k|^2 
- \frac{1}{\sigma^2}\sum_{k<l} |T_{kl}|^2\right)\,.
\end{equation} 
The integrals
on all  the eigenvalues will be discussed
in the next section. In the present
section we show how to perform
the integrals over $T_{kl}$.
To this end the notation $\langle \cdots \rangle_T$
is introduced, denoting a normalized integral on all
$T_{kl}$ with weight $\exp(-\sigma^{-2}\sum_{k < l} |T_{kl}|^2)$.

Consider first the average $\langle O_{11}\rangle_T$.
Let
\begin{equation}
S_l = \sum_{p=1}^l\langle |b_p|^2\rangle_T
\end{equation}
so that $S_1 = 1$ and $S_N = \langle O_{11}\rangle_T$. 
Then from Eq. (\ref{eq:recursion1})
\begin{equation}
\langle |b_l|^2\rangle_T = 
\frac{\sigma^2}{|\lambda_1-\lambda_l|^2}\,S_{l-1}\,,
\end{equation}
and hence
\begin{equation}
S_l = \Big(1+\frac{\sigma^2}{|\lambda_1-\lambda_l|^2}\Big)\,S_{l-1}\,.
\end{equation}
Together with $S_1=1$ this implies
 \begin{equation}
\label{eq:O11avT}
\langle O_{11}\rangle_T
= \prod_{l=2}^N\left(1+\frac{\sigma^2}{|\lambda_1-\lambda_l|^2}\right)\,.
\end{equation}

Consider now the average $\langle O_{12}\rangle_T$.
Let
\begin{equation}
S_l' = \langle \overline{T}_{12} \sum_{k=1}^l b_k \overline{d}_k\rangle_T
\end{equation}
so that $S_1' = 0$, $S_2' = \sigma^2/(\lambda_1-
\lambda_2)$ and
$\langle O_{12}\rangle_T= -S_N^\prime/(\overline{\lambda}_1-\overline{\lambda}_2)$.
Now
\begin{eqnarray}
S_{l+1}^\prime-S_l^\prime &= &\nonumber
\langle \overline{T}_{12}\, b_{l+1}\,\overline{d}_{l+1}\rangle_{T}\\
&=& \Big\langle \overline{T}_{12}\left[\nonumber
\frac{1}{\lambda_1-\lambda_{l+1}}
\sum_{q=1}^l b_l T_{ql+1}\right]
\left[\frac{1}{\overline{\lambda}_2-\overline{\lambda}_{l+1}}\sum_{k=1}^l
\overline{d}_k \overline{T}_{kl}\right]
\Big\rangle_T\\
&=&
\frac{\sigma^2}{(\lambda_1-\lambda_{l+1})
(\overline{\lambda}_2-\overline{\lambda}_{l+1})}\,S_{l}^\prime\,.
\end{eqnarray}
This implies
\begin{equation}
\label{eq:O12avT}
\langle O_{12}\rangle_T 
= -\frac{\sigma^2}{|\lambda_1-\lambda_2|^2} \prod_{l=3}^N
\left(1+\frac{\sigma^2}{(\lambda_1-\lambda_l)
        (\overline{\lambda}_2-\overline{\lambda}_l)}\right)\,.
\end{equation}
Eqs. (\ref{eq:O11avT}) and (\ref{eq:O12avT})
represent the averages of $O_{11}$ and $O_{12}$ 
with respect to the coordinates 
$T_{kl}$. The remaining integrals are those
over $\lambda_k$. Using Eqs. (\ref{eq:O11avT}) and (\ref{eq:O12avT})
one has
\begin{eqnarray}
\label{eq:P}
O(z_1) &=&N\sigma^2\,\Big\langle \delta(z_1-\lambda_1)
\prod_{l=2}^N\Big(1+\frac{\sigma^2}{|\lambda_1-\lambda_l|^2}\Big)\Big\rangle_P
\end{eqnarray}
and
\begin{eqnarray}
\label{eq:Q}
O(z_1,z_2) &=& -N(N-1)\,\sigma^2\,
\Big\langle \delta(z_1-\lambda_1)\,\delta(z_2-\lambda_2)\\
&\times& 
\frac{\sigma^2}{|\lambda_1-\lambda_2|^2}\,
\prod_{l=3}^N \Big(1+
\frac{\sigma^2}{(\lambda_1-\lambda_l)(\bar{\lambda}_2-\bar{\lambda}_l)}\Big)
\Big\rangle_P\,,
\nonumber
\end{eqnarray}
where $\langle \cdots \rangle_P$ is an average
with the weight (\ref{eq:pz}).

\subsubsection{The case $N=2$}
The case $N=2$ is particularly
simple. We find
\begin{equation}
\label{eq:O11N2}
O(z) = \frac{1}{\pi} \,\Big(2+\frac{\gap|z|^2}{\gap\sigma^2} \Big)\, 
\exp\Big(-\frac{\gap|z|^2}{\gap\sigma^2}\Big)
\end{equation}
and
\begin{equation}
\label{eq:O12N2}
O(z_1,z_2) 
= -\frac{1}{\sigma^2\pi^2} \,
\exp\Big(\!-\frac{\gap|z_1|^2}{\gap\sigma^2}
         -\frac{\gap|z_2|^2}{\gap\sigma^2}\Big)\,.
\end{equation}
These expressions are useful as simple checks
of results for arbitrary values of $N$.

Of more general interest, the distribution 
of $O_{\alpha\alpha}$ is,
from Eqs. (\ref{eq:recursion1}) and (\ref{eq:O11})
\begin{equation}
\label{eq:dist}
P(O_{\alpha\alpha})
= 4\,\frac{\Theta(O_{\alpha\alpha}-1)}{(2\,O_{\alpha\alpha}-1)^3}\,,
\end{equation}
where $\Theta(x)=1$ for $x>0$ and zero otherwise. This gives
\begin{equation}
\langle O_{\alpha\alpha} \rangle = 3/2
\end{equation}
which is consistent with 
(\ref{eq:O11N2}) integrated over $z$. Note that the second 
and higher moments of $O_{\alpha\alpha}$ diverge. 
We argue in section \ref{sec:res_dist}
that the 
the tail of the distribution of $O_{\alpha\alpha}$
at large $O_{\alpha\alpha}$ has the same
form for all $N \geq 2$. 

\subsubsection{Calculation of the eigenvalue averages}
\label{sec:lambda_int2}
In this section we show how to
evaluate the remaining integrals in (\ref{eq:P})
and (\ref{eq:Q}). They can be performed in the
same way as those in section \ref{sec:dosginibre}.
In analogy with Eq. (\ref{eq:detdos})
one has
\begin{equation}
\label{eq:detO11}
O(z) = N\sigma^2\, (\pi \,N!\,\sigma^2)^{-1}\,\,
\exp\Big(\!\!-\frac{\gap|z|^2}{\gap\sigma^2}\Big)\,
\det\left [
\begin{array}{ccccr}
g_{00} & g_{01} &  & &   0                      \\
g_{10} &        &\ddots  & &                    \\
 &\hspace*{-3mm}\ddots        &  & &            \\
       &  &  &            & g_{N-3N-2}          \\
0      &        &  & g_{N-2N-3} & g_{N-2N-2}   
\end{array}
\right]
\end{equation}
with $g_{ij} = (\pi\,j!\,\sigma^{2j+4})^{-1} \int\!d^2\lambda\, 
\overline{\lambda}^i{\rule[0mm]{0mm}{4mm}\lambda}^j \,
(\sigma^2+|z-\lambda|^2)\,\exp(-\sigma^{-2}|\lambda|^2)$.
Eq. (\ref{eq:detO11}) provides an explicit expression
for $O(z)$ 
for general $N$.
The determinant can be easily evaluated numerically,
as is shown in section \ref{sec:resultsnp}.
For $z=0$, the
$(N\!-\!1)\times (N\!-\!1)$ determinant
in Eq. (\ref{eq:detdos}) 
is simply diagonal. Denoting it by $G_{N-1}$ we have 
$G_{k+1} = (k+2) G_k$ and thus
\begin{equation}
\label{eq:resO11z0}
O(z) \,
\mbox{\rule[-4mm]{0.25pt}{8mm}\raisebox{-3mm}{\hspace{3pt}$z\!=\!0$}}\,\,
= \frac{N}{\pi}\,,
\end{equation}
independent of $\sigma^2$.
For $N=2$ this expression gives $O(0)=2/\pi$, consistent 
with Eq. (\ref{eq:O11N2}).
An expression for $O(z_1,z_2)$ can be obtained in analogy with
(\ref{eq:detR2}):
\begin{eqnarray}
\label{eq:detO12}
O(z_1,z_2) &= &
-N\sigma^2\,(\pi^2\,N!\,\sigma^4)^{-1}\,
\exp\Big(\!\!-\frac{\gap|z_1|^2}{\gap\sigma^2}
     -\frac{\gap|z_2|^2}{\gap\sigma^2}\Big)\\[0.5cm]
&\times &
\det\left [
\begin{array}{cccccr}
h_{00} & h_{01} &h_{02} &          &            &   0        \\
h_{10} & h_{11} &       & \ddots   &            &            \\
h_{20} &  \phantom{\ddots}      &       &          &            &            \\
       & \ddots &       &          &            & h_{N-5N-3} \\
       &        &       &          & h_{N-4N-4} & h_{N-4N-3} \\
0      &        &       &h_{N-3N-5}& h_{N-3N-4} & h_{N-3N-3}   
\end{array}
\right]
\nonumber
\end{eqnarray}
with 
\begin{eqnarray}
h_{ij} &=& (\pi\, (j\!+\!1)!\,\sigma^{2j+6})^{-1} 
\int\!d^2\lambda\, 
\overline{\lambda}^{i}{\rule[0mm]{0mm}{4mm}\lambda}^{j} \,
\Big[|z_1-\lambda|^2\,|z_2-\lambda|^2\\
&&\hspace*{2cm}+\sigma^2
(\overline{z}_1-\overline{\lambda}\big)
(z_2-\lambda) \Big]\,\exp\Big(\!\!-\frac{\gap|\lambda|^2}{\gap\sigma^2}\Big)\,.
\nonumber
\end{eqnarray}
For $z_2=0$ and $z_1=z$, and denoting
the determinant in (\ref{eq:detO12}) by
$H_{N-2}$,
we  obtain the recursion relation
\begin{equation}
H_{k+1} = (\sigma^{-2}\,|z|^2+k+3) H_k 
-\sigma^{-2}\,|z|^2\,(k+2)\, H_{k-1}\,.
\end{equation}
With $H_1=3+\sigma^{-2}\,|z|^2$ and $H_2 = 12
+4\,\sigma^{-2}\,|z|^2 +\sigma^{-4}\,|z|^4$ this yields
\begin{equation}
\label{eq:resO12z0}
O(z,0)=-N\sigma^2\pi^{-2}\,\exp\Big(\!\!-\frac{\gap|z|^2}{\gap\sigma^2}\Big)\,
\frac{1}{|z|^4}
\sum_{l=2}^{N} 
\frac{|z|^{2l}}{l!\,\sigma^{2l}}\,.
\end{equation}
For $N=2$ this gives 
$O(z,0)=-(\sigma\pi)^{-2}\,\exp(-\sigma^{-2}\,|z|^2)$, which is consistent
with (\ref{eq:O12N2}). An additional check is provided by
the fact that Eqs. (\ref{eq:resO11z0}) 
and (\ref{eq:resO12z0}) obey the sum rule (\ref{eq:sumruleD}).
In the limit of $N$ large, with $\sigma^2=N^{-1}$, we obtain,
for $z\neq 0$,
\begin{equation}
\label{eq:respert}
O(z,0) = -\frac{1}{\pi^2|z|^4}
\end{equation}
for $|z| < 1$ and zero otherwise.
In order to exhibit the behavior of
Eq. (\ref{eq:resO12z0}) near the origin,
for $\sigma^2=N^{-1}$ and in the large $N$
limit, we write $\omega = N^{1/2} z$; for 
$|\omega| \ll N^{1/2}$ we then have
\begin{equation}
\label{eq:resreg}
N^{-2}O(z,0) =  -\frac{1}{\pi^{2} |\omega|^4} \,\Big[
1-(1+|\omega|^2)\,{\rm e}^{-|\omega|^2}\Big]\,.
\end{equation}
Eq. (\ref{eq:resreg}) displays the way in which the result (\ref{eq:respert})
is regularized as $|z|\rightarrow 0$.

\subsubsection{Simplified calculation of the eigenvalue averages
               for $N$ large}
\label{sec:lambda_int3}
The main results of section \ref{sec:lambda_int2} are the determinantal
expressions Eqs. (\ref{eq:detO11})
and (\ref{eq:detO12}), providing exact results
for the eigenvector
correlators (\ref{eq:Odiag}) and (\ref{eq:Ooff}).
In the present
section we provide approximate expressions
for (\ref{eq:detO11}) and (\ref{eq:detO12}) which,
for $\sigma^2 = N^{-1}$,
are valid in the limit of $N$ large, with $z_1\neq z_2$
and   $|z_1|,|z_2| < 1$. 
 In the following we shall need to indicate explicitly the rank of
 the random matrix considered, and so we use the notation
 \begin{eqnarray}
O_N(z) &=& \Big\langle\frac{1}{N} \sum_\alpha O_{\alpha\alpha}
\,\delta(z-\lambda_\alpha)\Big\rangle\,,\\
O_N(z_1,z_2) &=& \Big\langle \frac{1}{N}
\sum_{\alpha\neq\beta} O_{\alpha\beta}
\,\delta(z_1-\lambda_\alpha) \,\delta(z_2-\lambda_\beta)
\Big\rangle\,.
\end{eqnarray}
in place of $O(z)$ and $O(z_1,z_2)$ 
[Eqs. (\ref{eq:Odiag}) and (\ref{eq:Ooff}) ].
Consider first $O_N(z)$.  We write
\begin{equation}
\label{eq:factorize}
O_N(z) \simeq  O_M(0)\, \Omega_1
\end{equation}
where
\begin{equation}
\Omega_1 = 
\prod_{l=M+1}^N 
\Big(1+\frac{1}{N}\frac{1}{|z-\lambda_l|^2}\Big)
\end{equation} 
and the product excludes the $M$ eigenvalues $\lambda_l$
closest in the complex plane to the point $z$,
as illustrated in Fig 1. We believe
that Eq. (\ref{eq:factorize}) is exact for $N\rightarrow\infty$
followed by $M\rightarrow\infty$, because we expect that
$\Omega_1$ has no fluctuations in that limit. 
This implies in particular
that we can calculate $\Omega_1$ by 
evaluating the average of its logarithm.
Starting from
\begin{equation}
\log \Omega_1 = \sum_{l=M+1}^N \log\Big(1+\frac{1}{N|z-\lambda_l|^2}\Big)
\end{equation}
and expanding the logarithm on the right side,
we have
\begin{equation}
\label{eq:1}
\log \Omega_1 = \frac{1}{N}\sum_{l=M+1}^N \frac{1}{|z-\lambda_l|^2} =
\int_{\cal D}\!
{\rm d}^2\lambda\, d(\lambda) \, \frac{1}{|z-\lambda|^2}
\end{equation}
where, in the large $N$ limit, $d(z)=\pi^{-1}$ 
for $|z|<1$ and $d(z)=0$ otherwise.
The domain $D$ of integration
excludes a disk of radius $\varrho$ with centred on $z$.
Since this disk should contain $M$ eigenvalues,
$\varrho^2 = M/N$. Thus we obtain
in the large $N$ limit
\begin{equation}
\Omega_1 = \varrho^{-2}\,(1-|z|^2)\,.
\end{equation}
Making use of the fact that $O_M(0) = M/\pi$
[see Eq. (\ref{eq:resO11z0})], and using
Eq. (\ref{eq:factorize}) we thus obtain 
\begin{equation}
\label{eq:resO11ca}
O(z) \simeq \frac{N}{\pi}(1-|z|^2)\,.
\end{equation}

\begin{figure}
  \vspace*{10mm}
  \centerline{\psfig{file=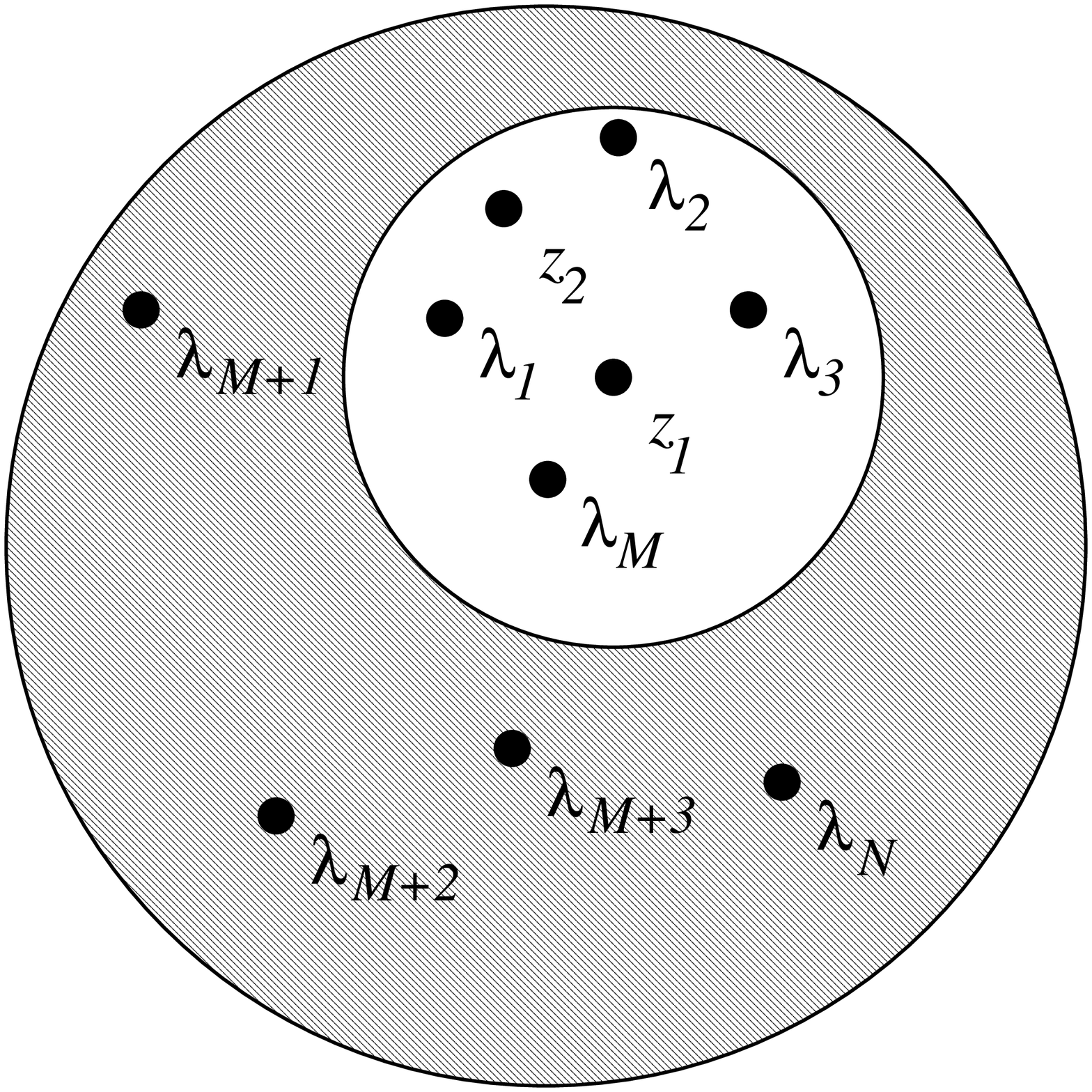,width=10cm}}
 \vspace*{10mm}
  \caption{\label{fig:disk} 
  Eigenvalue distribution in the complex plane, as discussed in Sec. \protect\ref{sec:lambda_int3}. }
 \end{figure}

The quantity $O_N(z_1,z_2)$ can
be calculated in a similar fashion.
To this end we write
\begin{equation}
\label{eq:f2}
O_N(z_1,z_2) = O_M(z_1-z_2,0) \Omega_2
\end{equation}
where $\Omega_2$ is 
\begin{equation}
 \Omega_2 =
  \prod_{l=M+1}^N 
 \Big(1+\frac{1}{N}
 \frac{1}{(z_1-\lambda_l)(\overline{z}_2-\overline{\lambda}_l)}\Big)
 \end{equation}
and the product excludes
the $M$ eigenvalues closest to $z_2$, with $1 \ll M \ll N$.
We first consider the case $|z_1-z_2| \ll 1$ .
Proceeding as above, we have 
\begin{equation}
\log \Omega_2 = 
\int_{\cal D}\!{\rm d}^2 \lambda\,d(\lambda)\,
\frac{1}{(z_1-\lambda)(\overline{z}_2-\overline{\lambda})}
\end{equation}
where again the domain of integration $\cal D$ 
is the unit disk with a disk of radius $\varrho$ around
$z_2$ removed as illustrated in Fig. \ref{fig:disk}.
In the large $N$ limit we obtain
\begin{equation}
\Omega_2 \simeq \varrho^{-2}(1-z_1\overline{z}_2)\,.
\end{equation}
As before, $\varrho^2 = M/N$. Using Eqs. (\ref{eq:resO12z0})
and (\ref{eq:f2}), we find in the large $N$ limit 
and with $\omega \equiv N^{1/2}\,(z_1-z_2)$, $|\omega| \ll {N}^{1/2}$
\begin{equation}
\label{eq:resO12ca}
N^{-2}O(z_1,z_2) \simeq -\,\frac{1-z_1\overline{z}_2}{\pi^2|\omega|^4}\,
\big(1-(1+|\omega|^2)\, {\rm e}^{-|\omega|^2}\big)\,.
\end{equation}
Second we consider the case $|z_1-z_2| \gg N^{-1/2}$.
In this case we obtain, in the large $N$ limit,
for $z_1\neq z_2$ and $|z_1|,|z_2| < 1$,
\begin{equation}
\label{eq:O12ca}
O(z_1,z_2) = -\frac{1}{\pi^2}\frac{1-z_1\overline{z}_2}{|z_1-z_2|^4}\,.
\end{equation}
For $|z_1|,|z_2| > 1$, $O(z_1,z_2)$ vanishes in this
limit.

\section{Girko's ensemble}
\label{sec:p}
\label{sec:girko}
In this section, we present a general approach
to calculating the averages of Eqs.
(\ref{eq:Odiag}) and (\ref{eq:Ooff})
perturbatively, using an expansion in powers of $N^{-1}$.
This will enable
us to treat more general ensembles
than the one considered
in the previous section. As an example, expressions
are derived for the averages of Eqs. 
(\ref{eq:Odiag}) and (\ref{eq:Ooff})
in the case of
Girko's ensemble, defined in Eq. (\ref{eq:girko}).
The expressions derived below
are appropriate for large $N$
and $z_1 \neq z_2$ in (\ref{eq:Ooff}).
For $\tau = 0$, Eqs. (\ref{eq:O12ca}) and (\ref{eq:resO11ca})
are thus reproduced. 
In the following
we set $\sigma = N^{-1/2}$:
the results
derived are correct in the large $N$ limit.

\subsection{Self-consistent Born approximation}
\label{sec:scba}
The desired approximations for (\ref{eq:Odiag})
and (\ref{eq:Ooff}) 
are obtained by calculating the average  in Eq. (\ref{eq:densD})
using Green functions.
The corresponding Green functions are non-analytic within
the support of the density of states
which occupies a finite region in the complex plane. 
In general, perturbation theory
yields only the analytic contribution, and in conventional
problems singularities on the real axis are obtained
by analytic continuation. In the present case one thus
proceeds as follows.
A Hermitian $2N\times 2N$ matrix $\bbox{H} = \bbox{H}_0 + \bbox{H}_1$
is introduced  \cite{cha97,efe97a,jan97,zee97,jan97b,jan97c}
\begin{equation}
\label{eq:H}
\bbox{H}_0 = 
\left(
\begin{array}{cc}
\eta \,\,& \\  & -\eta
\end{array}
\right )\,,\hspace*{5mm}
\bbox{H}_1 = 
\left(
\begin{array}{cc}
 \,\,& A\\ A^\dagger &
\end{array}
\right )
\end{equation}
with $\eta > 0$, $A=z-J$ and with inverse
\begin{equation}
\label{eq:G}
\bbox{G}
=
\left( 
\begin{array}{cc}
\eta[\eta^2+ A A^\dagger]^{-1} & A[\eta^2+ A^\dagger A]^{-1}\\
       A^\dagger[\eta^2+ A A^\dagger]^{-1} & -\eta[\eta^2+ A^\dagger A]^{-1}
\end{array}
\right)
\equiv
\left( 
\begin{array}{cc}
G_{11}& G_{12}\\
G_{21}& G_{22}
\end{array}
\right)\,.
\end{equation}

\begin{figure}
  \vspace*{1cm}
  \hspace*{1.4cm} {\large (a)} Diagrammatic notation for the 
    ensemble (\ref{eq:girko}):

  \vspace*{1cm}
  \hspace*{1.4cm}
  $\langle G_{11}(z)\rangle%
   =\hbox{\protect\raisebox{-1mm}
         {\protect\psfig{file=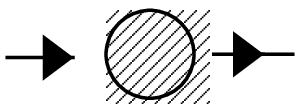,width=1cm}}}$\,, 
    \hspace*{1cm}
  $\langle G_{21}(z)\rangle%
   =\hbox{\protect\raisebox{-1mm}
         {\protect\psfig{file=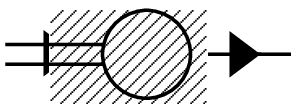,width=1cm}}}$\,,

  \vspace*{1.0cm}
  \hspace*{1.4cm}
  $\langle G_{12}(z)\rangle%
   =\hbox{\protect\raisebox{-1mm}
         {\protect\psfig{file=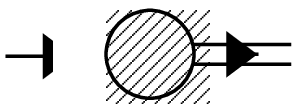,width=1cm}}}$\,,
    \hspace*{1cm}
  $\langle G_{22}(z)\rangle%
   =\hbox{\protect\raisebox{-1mm}
         {\protect\psfig{file=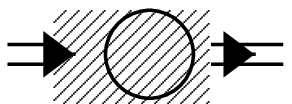,width=1cm}}}$\,,

   \vspace*{0.5cm} 
   \hspace*{1.4cm}
   $\langle J_{kl} \overline{J}_{kl}\rangle 
     =\hbox{\protect\raisebox{-4mm}
          {\protect\psfig{file=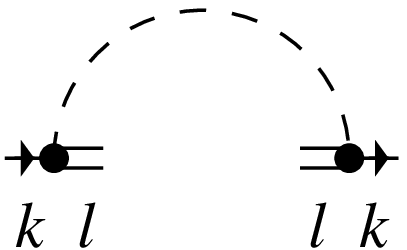,width=2cm}}}$\,,
   \hspace*{0.15cm}
   $\langle J_{kl} {J}_{lk}\rangle 
     =\hbox{\protect\raisebox{-4mm}
          {\protect\psfig{file=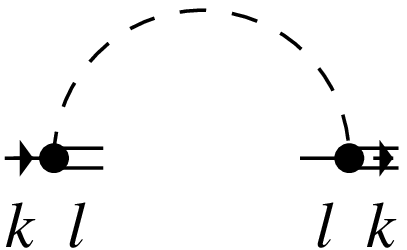,width=2cm}}}$\,,\hspace*{0.2cm}
  $z =  -\,\hbox{\protect\raisebox{-0.1mm}
       {\protect\psfig{file=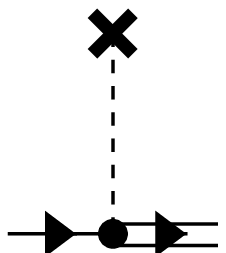,width=0.8cm}}}$\,,\hspace*{0.2cm}
  $\overline{z} =%
        -\,\hbox{\protect\raisebox{-0.1mm}
       {\protect\psfig{file=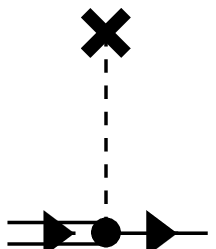,width=0.8cm}}}$\,.
  \vspace*{0.5cm}

  \hspace*{1.4cm} {\large (b)} Self energy $\bbox{\Sigma}$:\\

  \vspace*{0.2cm}
  \hspace*{1.4cm}
  {\psfig{file=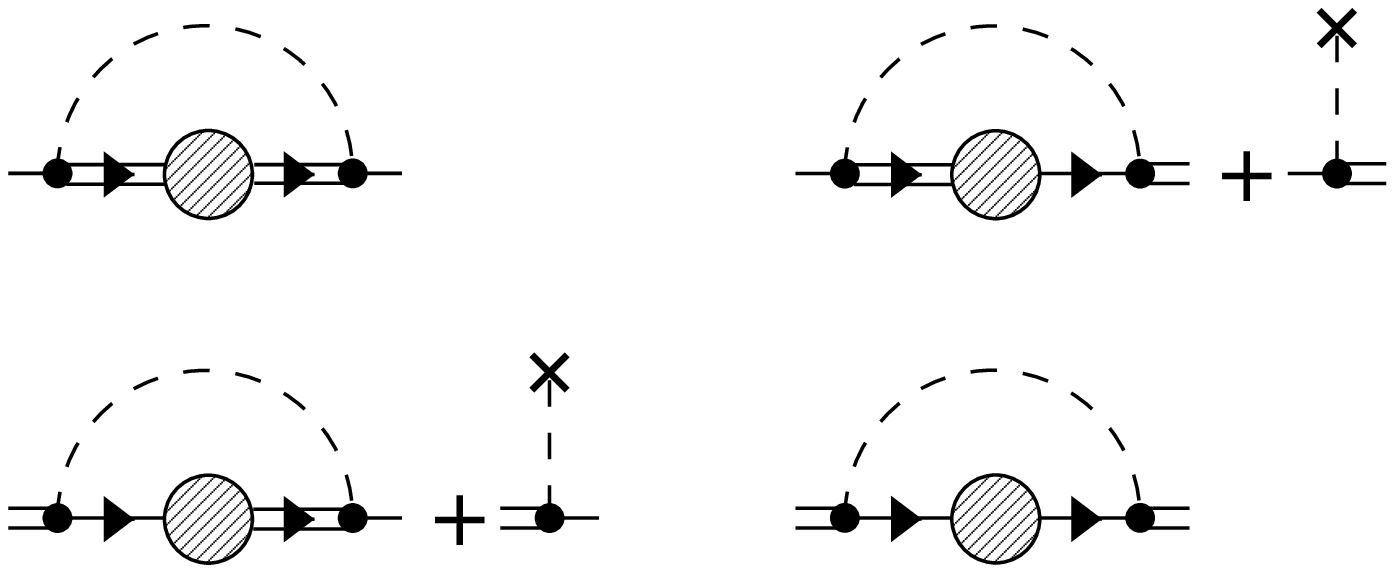,width=6cm}}
  \vspace*{1.0cm}
  \caption{\label{fig:dyson}%
  (a) Diagrammatic notation for the ensemble (\ref{eq:girko}). (b)
  Self-energy $\bbox{\Sigma}$ in equation (\ref{eq:sc}), 
  to lowest order in $N^{-1}$,
  compare Eq. (\ref{eq:se}).
  Note that a summation over internal
  indices in closed loops incurs an additional factor of $N$.}
\end{figure}

Expanding the Green function as a power series
in $\bbox{H}_1$, its ensemble average $\langle \bbox{G}\rangle$ 
can be written as
\begin{equation}
\label{eq:sc}
\langle\bbox{G}\rangle = \bbox{G}_0 + \bbox{G}_0\bbox{\Sigma}
\langle \bbox{G}\rangle\,,
\end{equation}
where $\bbox{G}_0 = \bbox{H}_0^{-1}$ and $\bbox{\Sigma}$
is a self-energy. Within the self-consistent
Born approximation one obtains
\cite{cha97,jan97}
\begin{equation}
\label{eq:se}
\bbox{\Sigma}
= \bone_N \otimes
\left (
\begin{array}{cc}
\langle G_{22}\rangle & -z+\tau \langle G_{21}\rangle\\
-\bar{z}+\tau \langle G_{12}\rangle& \langle G_{11}\rangle
\end{array}
\right)
\end{equation}
as illustrated diagrammatically in Fig. \ref{fig:dyson}.
The self-consistent Born approximation is exact
in the limit  $N\rightarrow \infty$. 
For $\eta\rightarrow 0$, the self-consistent solution 
of Eqs. (\ref{eq:sc}) and (\ref{eq:se}) is as
follows \cite{som88,cha97,jan97}:
one has for all $z$
$\langle G_{22}\rangle = - \langle G_{11}\rangle$ and
$\langle G_{12} \rangle = \langle \overline{G}_{21}\rangle$.
In addition, $\langle G_{11}\rangle$ is non-zero
only inside the ellipse defined
by $[x/(1+\tau)]^2+[y/(1-\tau)]^2 = 1$
\begin{eqnarray}
\left\langle G_{11}\right\rangle
&=&
\left\{
\begin{array}{ll}
\sqrt{1-[x/(1+\tau)]^2-[y/(1-\tau)]^2}
&\mbox{inside the ellipse}\,,\\
0 &\mbox{outside}\,.
\end{array}\right .
\end{eqnarray}
Furthermore,
\begin{eqnarray}
\label{eq:G21}
\left\langle G_{21}\right\rangle &=&
\left \{
\begin{array}{ll}
{x}/{(1+\tau)}-{{\rm i} y}/{(1-\tau)} 
&\hspace{1.85cm}\mbox{inside the ellipse}\,,\\
\left (z-\sqrt{z^2-4\tau}\right)/2\tau
&\hspace{1.85cm}\mbox{outside}\,.
\end{array}
\right .
\end{eqnarray}

 Using        Eq. (\ref{eq:dos}), the density
 of states is given by
 \begin{equation}
 d(z) = \lim_{\eta\rightarrow 0} 
\frac{1}{\pi}\frac{\partial}{\partial\bar{z}}
 \left\langle
 G_{21} \right\rangle\,.
 \end{equation}
It thus turns out that for $N\gg 1$ the support of
$d(z)$ is an ellipse in the complex plane
\cite{som88,cha97,jan97} with
\begin{equation}
\label{eq:dgirko}
d(z) = 
\left \{
\begin{array}{ll}
\pi^{-1}(1-\tau^2)^{-1} & \mbox{for $[x/(1+\tau)]^2+[y/(1-\tau)]^2 < 1$}\,,\\
0                       & \mbox{otherwise}\,.
\end{array}
\right .
\end{equation}
In the limit $\tau\rightarrow 1$, the eigenvalue density
of the Gaussian Unitary Ensemble
is recovered, for which
$ d(z) = \delta(y) \,(2\pi)^{-1} \sqrt{4-x^2}$.
Alternatively,
setting $\tau = 0$, the support of the density of
states in the complex plane becomes a disk of unit radius
centred around the origin [compare Eq. (\ref{eq:doslargeN})].

\subsection{Bethe-Salpeter equation}
\label{sec:bse}
\begin{figure}
\hspace*{1.4cm}\psfig{file=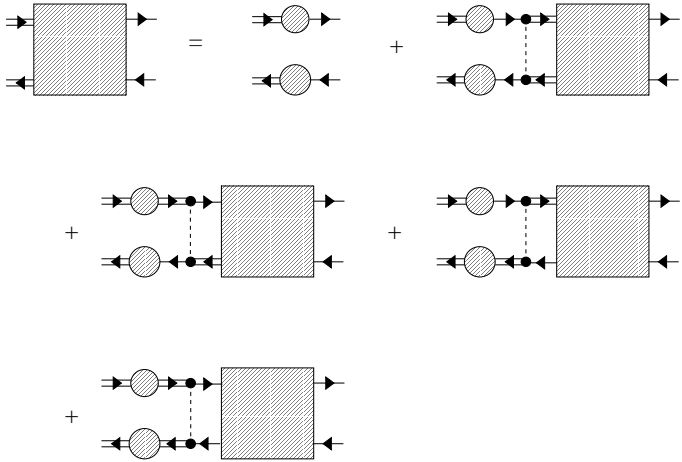,width=9cm}
\vspace*{1cm}
\caption{\label{fig:bse} Equation for the
matrix product $\langle G_{21}(1)\,G_{12}(2)\rangle$.
The diagrammatic rules are
as in Fig. \protect\ref{fig:dyson}.}
\end{figure}

In the following, $G_{kl}(z_1,\bar z_1)$
is denoted by $G_{kl}(1)$ ($k,l=1,2$).
An equation for
the average of the matrix product
$\langle G_{21}(1)\,G_{12}(2)\rangle$,
accurate at leading order in $N^{-1}$,                    
is shown  diagrammatically in Fig. \ref{fig:bse}. 
There are sixteen
such equations for all products
$\langle G_{ij}(1)\,G_{kl}(2)\rangle$ for
$i,\ldots,l=1,2$. In order to write these in
matrix form, one defines 
\begin{equation}
\bbox{R}(1,2) = \langle\bbox{G}(1)\otimes\bar{\bbox{G}}(2)\rangle\,.
\end{equation}
Similarly, $\bbox{R}_0(1,2)$ is the matrix
$\bra \bbox{G}(1)\ket \otimes \bra \overline{\bbox{G}}(2) \ket$. The
matrices $\bbox{R}$ and $\bbox{R}_0$ are Hermitian. 
Defining the vertex
\begin{equation}
\bbox{\Gamma} = \bone_N \otimes
\left(
\begin{array}{cccc}
&&&1\\
&&\tau&\\%\hline
&\tau&&\\
1&&&
\end{array}
\right)\,,
\end{equation}
the diagrammatic expression     for $\bbox{R}(1,2)$
can be written as
\begin{equation}
\bbox{R}(1,2) = \bbox{R}_0(1,2) + \bbox{R}_0(1,2)\,\bbox{\Gamma} \,
\bbox{R}(1,2)\,.
\end{equation}
Eq. (\ref{eq:bse})  has the solution
\begin{equation}
\label{eq:bse}
\bbox{R}(1,2) = \left[1-\bbox{R}_0(1,2)\,\bbox{\Gamma}\right]^{-1}\,
\bbox{R}_0(1,2)\,.
\end{equation}
We first discuss the simplest case, $\tau =0$. If
$z_1$ and $z_2$ lie inside the support of the density of states
($|z_1|<1$ and $|z_2|<1$)
\begin{equation}
\label{eq:R1}
\bbox{R}_0(1,2) = \bone_N \otimes
\left(
\begin{array}{cc}
\sqrt{1-z_1\bar{z}_1} & z_1\\
\bar{z}_1 & -\sqrt{1-z_1\bar{z}_1}
\end{array}
\right)
\otimes
\left(
\begin{array}{cc}
\sqrt{1-z_2\bar{z}_2} & \bar{z}_2\\
{z}_2 & -\sqrt{1-z_2\bar{z}_2}
\end{array}
\right)\,.
\end{equation}
In this case, from Eq. (\ref{eq:bse})
\begin{equation}
\bbox{R}(1,2)\!=\!\bone_N\!\otimes\!
\left(
\begin{array}{cccc}
\rule[-4mm]{0mm}{11.5mm}
{\sqrt{\ds 1-z_1\bar{z}_1}\sqrt{\ds 1-z_2\bar{z}_2}\over{\ds |z_1-z_2|^2}}&
-{\sqrt{\ds 1-z_1 \bar{z}_1}\over{\ds z_1-z_2}}\;&\;
 \sqrt{\ds 1-z_1 \bar{z}_1}\over{\ds\bar{z}_1-\bar{z}_2}&
\ds 1-z_1\bar{z}_1-z_2\bar{z}_2+z_1\bar{z}_2\over{\ds |z_1-z_2|^2} 
\\\rule[-4mm]{0mm}{11.5mm}
-{{\sqrt{\ds 1-z_1 \bar{z}_1}}\over{\ds \bar{z}_1-\bar{z}_2}}&
0&
-{{\ds z_1-z_2}\over{\ds \bar{z}_1-\bar{z}_2}}&
-{{\sqrt{\ds 1-z_2 \bar{z}_2}}\over{\ds \bar{z}_1-{z}_2}}
\\%\hline
\rule[-4mm]{0mm}{11.5mm}
{\sqrt{\ds 1-z_2 \bar{z}_2}\over{\ds z_1-z_2}} &
-{{\ds \bar{z}_1-\bar{z}_2}\over{\ds z_1-z_2}}&
0&
{\sqrt{\ds 1-z_1 \bar{z}_1}\over{\ds z_1-z_2}}
\\\rule[-4mm]{0mm}{11.5mm}
{\ds 1-z_1\bar{z}_1-z_2\bar{z}_2+\bar{z}_1z_2\over{\ds |z_1-z_2|^2}}&
-{\sqrt{\ds 1-z_2 \bar{z}_2}\over{\ds z_1-z_2}}&
{\sqrt{\ds 1-z_1 \bar{z}_1}\over{\ds \bar{z}_1-\bar{z}_2}}&
{\sqrt{\ds 1-z_1\bar{z}_1}\sqrt{\ds 1-z_2\bar{z}_2}\over{\ds |z_1-z_2|^2}}
\end{array}
\right)\,.
\end{equation}
Alternatively, if both $|z_1|>1$ and $|z_2|>1$, we obtain   
\begin{equation}
\label{eq:R2}
\bbox{R}(1,2) = \bone_N\otimes
\left(
\begin{array}{cccc}
&&&\ds 1\over{\ds \bar{z}_1 z_2 -1}\\ 
&&\ds 1\over{\ds \bar{z}_1\bar{z}_2} &\\%\hline
&\ds 1\over{\ds z_1 z_2} &&\\
\ds 1\over{\ds z_1 \bar{z}_2 -1}&&&
\end{array}\right)\,.
\end{equation}

The general case, $\tau\neq 0$, is dealt with as follows. 
We define a transformation
\begin{equation}
\label{eq:map}
z \rightarrow w = {z-\tau\bar{z}\over 1-\tau^2}
\end{equation}
which maps the support of the density of states in the $z$-plane
onto the unit disk in the $w$-plane. For $z_1$ and $z_2$ inside
the support of the density of states
one has $|w_1|<1$, $|w_2|< 1$ and 
\begin{equation}
\bbox{R}_0(1,2) = \bone_N\otimes
\left(
\begin{array}{cc}
\sqrt{1-w_1\bar{w}_1} & w_1\\
\bar{w}_1 & -\sqrt{1-w_1\bar{w}_1}
\end{array}
\right)
\otimes
\left(
\begin{array}{cc}
\sqrt{1-w_2\bar{w}_2} & \bar{w}_2\\
{w}_2 & -\sqrt{1-w_2\bar{w}_2}
\end{array}
\right)\,.
\end{equation}
The resulting matrix $\bbox{R}(1,2)$ is more complicated
than Eq. (\ref{eq:R1}). For the element
$\langle {G}_{21}(1) {G}_{12}(2)\rangle $
in the case $|w_1|<1$ and $|w_2|< 1$
we find
\begin{eqnarray}
\label{eq:result}
\lefteqn{\langle {G}_{21}(1) {G}_{12}(2)\rangle}\\
\nonumber
 &=& {
(1-\tau)^2 +(1+\tau^2) \bar{w}_1w_2 -w_1\bar{w}_1-w_2\bar{w}_2
+\tau(\bar{w}_1 \bar{w}_2 +w_1 w_2  -\bar{w}_1^2 -w_2^2)
\over |w_1-w_2+\tau(\bar{w}_1-\bar{w}_2)|^2}\,.
\end{eqnarray}

\subsection{Calculation of the density $D(z_1,z_2)$}
\label{sec:resultp}
\label{sec:D}
The density $D(z_1,z_2)$ 
can be expressed 
in terms of Green functions, from 
Eq. (\ref{eq:densD}), as
\begin{equation}
D(z_1,z_2) = \lim_{\eta\rightarrow 0}\frac{1}{\pi^2}
\frac{\partial}{\partial \bar{z}_1}
\frac{\partial}{\partial {z}_2}\,
\left\langle
 G_{21}(1) G_{12}(2)\right\rangle\,.
\end{equation}
We find from Eq. (\ref{eq:result}), for $|z_1|,|z_2|$ within the ellipse,
that
\begin{equation}
\label{eq:Dgirko}
D(z_1,z_2) = -\frac{(1-\tau)^2}{\pi^2}\, 
\frac{(1-\tau^2)^2 -(1+\tau^2)z_1\bar{z}_2
+\tau(z_1^2 + \bar{z}_2^2)}{|z_1-z_2|^4}\,.
\end{equation}
For $z_1$ and $z_2$ outside the  ellipse,
$D(z_1,z_2)$  vanishes. 

As a check it can be shown explicitly that
$D(z_1,z_2)$ obeys the sum rule
(\ref{eq:sumruleD}). Using
Green's theorem, we have
\begin{equation}
\int\!{\rm d}^2z_2\, D(z_1,z_2)
 = 
\frac{1}{\pi} \frac{\partial}{\partial \overline{z}_1}
\frac{1}{2\pi {\rm i}} 
\oint {\rm d}z_2\,
\langle G_{21}(1)G_{12}(2)\rangle
\end{equation}
where the contour integral is around the ellipse.
By means of the transformation (\ref{eq:map}), this contour
may be mapped into the unit circle in the $w$-plane, giving
\begin{eqnarray}
\int\!{\rm d}^2z_2\, D(z_1,z_2)
&=&
\frac{1}{\pi} \frac{1}{1\!-\!\tau^2}
\left(\frac{\partial}{\partial \overline{w}_1} 
-\tau\frac{\partial}{\partial w_1}\right)
\frac{1}{2\pi{\rm i}}
\oint 
\hspace{-0.85cm}\mbox{\raisebox{-7mm}{$\scriptstyle |w_2|=1$} }
({\rm d}w_2 + \tau {\rm d} \overline{w}_2)\,
\langle G_{21}(1)G_{12}(2)\rangle\nonumber\\
&=& \frac{1}{\pi}\frac{1}{1\!-\!\tau^2}
\end{eqnarray}
for $|w_1| < 1$ and zero otherwise, as expected from Eq. (\ref{eq:dgirko}).

As a final check we observe that, with $\tau = 0$,
Eq. (\ref{eq:Dgirko}) implies
\begin{equation}
\label{eq:ndiag}
O(z_1,z_2) =   -\frac{1}{\pi^2}\frac{1-z_1 \overline{z}_2}{|z_1-z_2|^4}
\end{equation}
for $|z_1|,|z_2| <1 $ and zero otherwise. 
Thus, our previous result, Eq. (\ref{eq:O12ca}), is reproduced from (\ref{eq:Dgirko})
for $\tau=0$. 

As pointed out in section \ref{sec:denslrev}, the
diagonal correlator $O(z)$ is given in terms
of the singular part of $D(z_1,z_2)$, see Eq. (\ref{eq:DOO}).
This singular part is inaccessible perturbatively,
in lowest order in $N^{-1}$\cite{note}. In order to determine
$O(z)$ within the perturbative approach discussed
in this section, we proceed as follows.
For simplicity, consider the case $\tau=0$.
Integrating the density $D(z_1,z_2)$ over a small disk
around $z_2$, of radius $\eta$ which is taken to
be small
\begin{equation}
\int\hspace{-0.85cm}\mbox{\raisebox{-6mm}{$\scriptstyle |z_1-z_2|\leq\eta$} }
\hspace*{-0.5cm}
d^2z_2\,D(z_1,z_2) =
\frac{1}{2\pi {\rm i}} \oint
\hspace{-0.85cm}\mbox{\raisebox{-6mm}{$\scriptstyle z_1-z_2=\eta$}}
\hspace*{-0.25cm}dz_2\,
\frac{1}{\pi}
\frac{\partial}{\partial \overline{z}_1}
\left\langle  G_{21}(1)\, G_{12}(2)\right\rangle
\simeq \frac{1}{\pi\eta^2} (1-|z_1|^2)\,,
\end{equation}
provided $z_1$ is sufficiently far away from the boundary.
On the other hand, from Eq. (\ref{eq:defD})
and for $\eta \simeq N^{-1/2}$, this
is approximately $O(z_1)$, so that up
to prefactors of order $O(1)$,
\begin{equation}
\label{eq:diag}
O(z_1) \simeq N (1-|z_1|^2)
\end{equation}
[compare Eq. (\ref{eq:resO11ca})]
and thus $O_{\alpha\alpha} \sim N$.
The sum rule (\ref{eq:sumruleD}) can be used to check
the consistency of Eqs. (\ref{eq:ndiag}) and (\ref{eq:diag}).

\section{Summary and discussion of the results}
\label{sec:resultsnp}
\label{sec:results}
In the present section we summarize
and discuss the results obtained in the previous
two sections. As in Sec. \ref{sec:girko},
the variance $\sigma^2$ in Eqs. (\ref{eq:ginibre})
and (\ref{eq:girko}) is taken to be $1/N$.

\subsection{Ginibre's ensemble}
\subsubsection{Eigenvector correlators Eqs. 
(\protect\ref{eq:Odiag}) and (\protect\ref{eq:Ooff})}
\label{sec:res_evcorr}
In the case of Ginibre's ensemble we have
been able to obtain exact expressions
for the eigenvector correlators,
Eqs. (\ref{eq:Odiag}) and (\ref{eq:Ooff}),
in the form of determinants. In certain
cases, we could simplify these expressions
further by recursion. Combining
these results [compare Eqs. (\ref{eq:resO11z0}) and
(\ref{eq:resO12z0})] with a continuum
treatment     (see section \ref{sec:lambda_int3}), 
in a way which we believe gives exact results
for the large $N$ limit, we have
for $|z_1- z_2| \neq 0$ and $|z_1|,|z_2| <1 $
\begin{eqnarray}
\label{eq:resO11oldp}
N^{-1}O(z_1)     &=& \frac{1}{\pi} (1-|z_1|^2)\\
\label{eq:resO12oldp}
O(z_1,z_2) &=& -\frac{1}{\pi^2} \frac{1-z_1\overline{z}_2}{|z_1-z_2|^2}
\end{eqnarray}
For $|z_1|, |z_2| \geq 1$, both densities
vanish as $N\rightarrow\infty$. To display
the form of $O(z_1,z_2)$ as $|z_1-z_2|\rightarrow 0$,
it is necessary to express $z_1-z_2$ in units
of the separation between adjacent eigenvalues.
Let $z_+ = (z_1+z_2)/2$,
$z_- = z_1-z_2$, and $\omega = \sqrt{N} z_-$.
For $|z_+|<1$, $\omega \ll \sqrt{N}$
and for  $N\gg 1$, Eq. (\ref{eq:resO12ca}) implies
\begin{equation}
\label{eq:singoldp}
N^{-2}O(z_1,z_2) = 
-\frac{1-|z_+|^2}{\pi^2|\omega|^4}
\,(1-(1+|\omega|)^2)\,{\rm e}^{-|\omega|^2})\,.
\end{equation}

\begin{figure}
  \centerline{\psfig{file=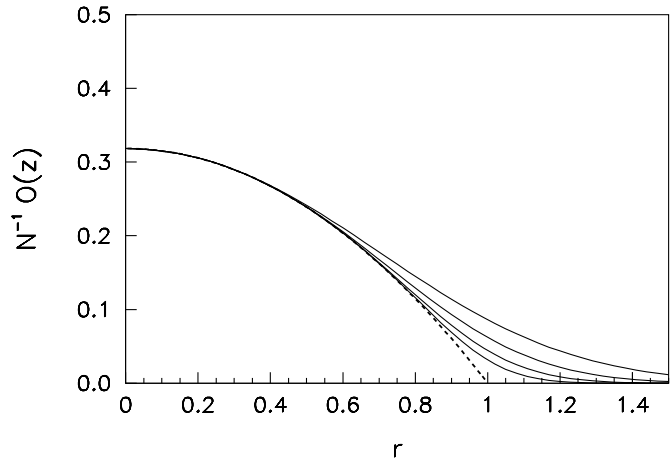,width=10cm}}
  \caption{\label{fig:diag}\protect\small
   $ N^{-1}\, O(z) =
  \langle N^{-2}\sum_\alpha O_{\alpha\alpha}\,\delta(z-\lambda_1)\rangle $
  as a function of $z = r\exp i\varphi$. The ensemble
  average is independent of $\varphi$. Results
  for $N=2,4,8,16$ are shown, together with
Eq. (\protect\ref{eq:resO11oldp}) 
  valid for large $N$ ($- - -$).}
\end{figure}

We have examined the convergence towards
these results for increasing $N$. In
Fig. \ref{fig:diag} we show $N^{-1}\,O(z)$ as a function
of $z$ for $N=2,4,8$ and $16$, obtained 
by evaluating the determinant in (\ref{eq:detO11}).
We also compare this with Eq. (\ref{eq:resO11oldp}).
The exact results converge
rapidly towards the approximate result (\ref{eq:resO11oldp})
as $N$ is increased,
provided $z$ is sufficiently
far from the boundary of the support of $O(z)$.

\begin{figure}
  \centerline{\psfig{file=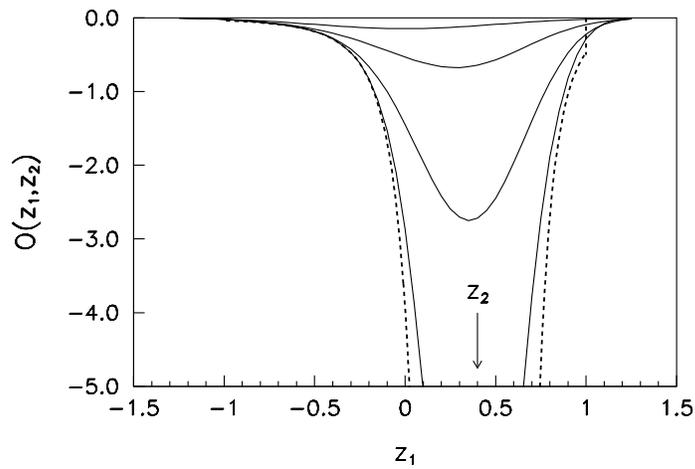,width=10cm}}
  \caption{\label{fig:ndiag}\protect\small
   $ O(z_1,z_2) =
  \langle N^{-1}\sum_{\alpha\neq\beta}
   O_{\alpha\beta}\,
   \delta(z_2-\lambda_\alpha)\,\delta(z_2-\lambda_\beta)\rangle $
   for $z_2 = 0.4$ as a function of $z_1$ on the real axis.
   Results
   for $N=2,4,8,16$ are shown, together with
   Eq. (\protect\ref{eq:resO12oldp}) 
   valid for large $N$ ($- - -$).}
\end{figure}

\begin{figure}
\centerline{\psfig{file=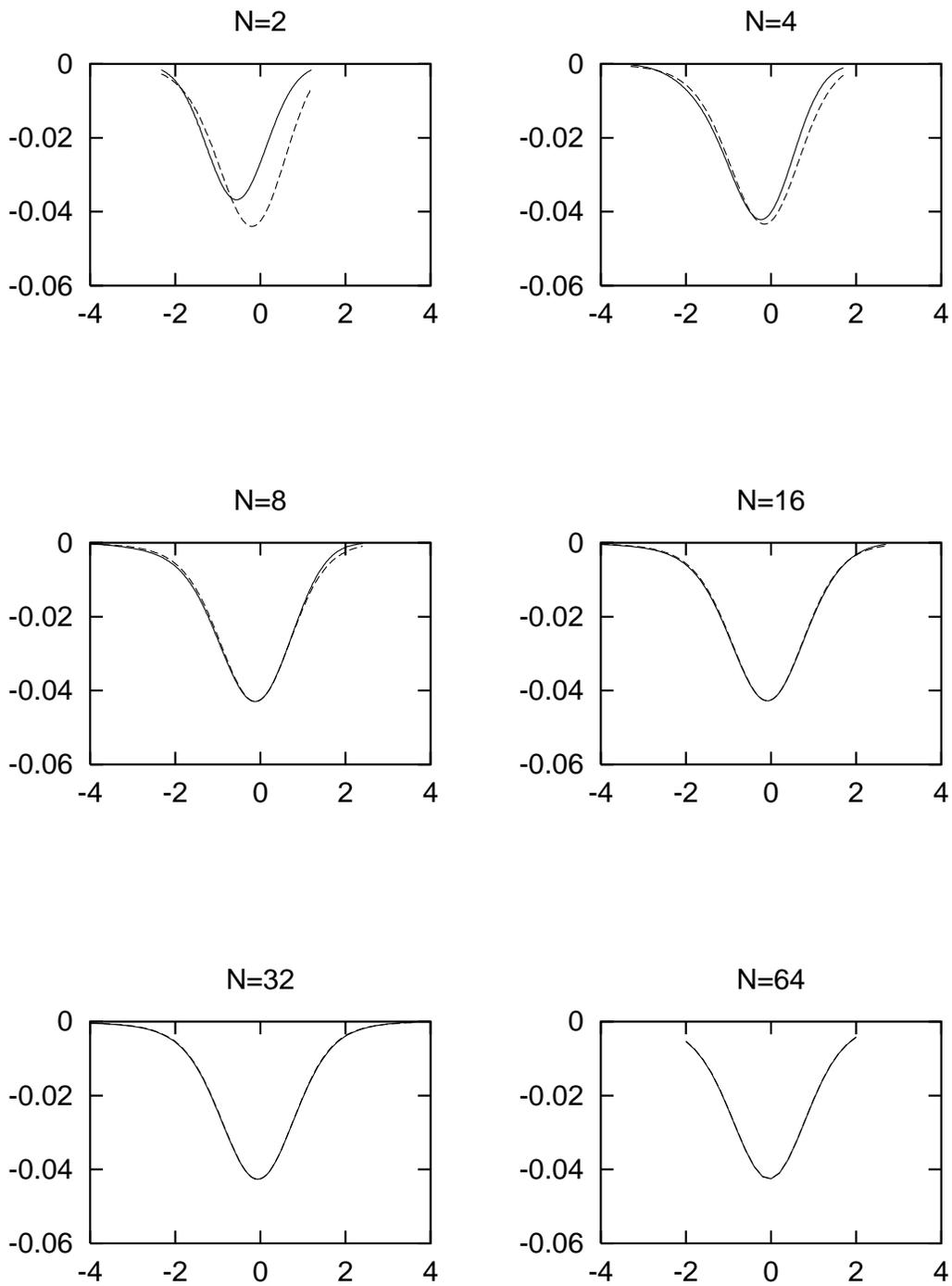,width=16cm}}
\vspace*{1cm}
\caption{\label{fig:sing}
Shows Eq. (\ref{eq:singoldp}) for $z_+ = 0.4 $ as
a function of $\omega$ for several values of $N$ 
($- - -$). Also shown are the corresponding exact results
($-\!\!\!-\!\!\!-\!\!\!-$).}
\end{figure}

In Fig. \ref{fig:ndiag} we show $O(z_1,z_2)$ as a function
of $z_1$ (on the real axis) for $z_2 = 0.4$ for
$N=2,4,8$ and $16$, obtained by evaluating
the determinant in (\ref{eq:detO12}). We compare this with 
Eq. (\ref{eq:resO12oldp}). 
Again, the exact results converge rapidly
towards the approximate expression (\ref{eq:resO12oldp})
as $N$ is increased, provided $z_1$ and $z_2$ are not
too close to each other or to the boundary of the 
support of $O(z_1,z_2)$.
Finally, in Fig. \ref{fig:sing} we show the behavior of     
$N^{-2}O(z_1,z_2)$ for  $|z_1-z_2| \stackrel{<}{\sim} N^{-1/2}$, 
comparing the approximate expression (\ref{eq:singoldp})
with exact results obtained by evaluating
the determinant in Eq. (\ref{eq:detO12}).
The exact results converge very rapidly to
the approximate expression as $N$ is increased,
provided $z_+ < 1$ and $|\omega | \ll \sqrt{N}$.

It is important to stress the dramatic difference between the behaviour of
 $O_{\alpha\beta}$ in Ginibre's ensemble and its behaviour 
in the case of Hermitian matrices, for which $O_{\alpha\beta}
=\delta_{\alpha\beta}$.
The fact that, by contrast, $O_{\alpha\alpha} \sim N$ in the non-Hermitian 
ensemble
can be understood as the behaviour
which would result if $\langle L_\alpha |$
and $| R_\alpha \rangle$ were independent random vectors, subject
to the normalisation of Eq.\,(\ref{eq:biorth}):
Choosing a basis and scaling in which
$|R_\alpha\rangle =
(1,0,\ldots,0)^T$, and assuming that
$\langle L_\alpha |$ is a random vector, biorthogonality
requires $\langle L_\alpha | = (1,b_2,\ldots,b_N)$,
where the coefficients $b_j$, for $j > 1$ are random
and $|b_j|$ is expected to be of order $O(1)$.
Thus $\langle O_{\alpha\alpha}\rangle \sim N$.
Moreover, large values 
for the diagonal elements of the matrix $O_{\alpha\beta}$ must 
be accompanied by some large (or many small)
off-diagonal elements, since the two are linked by 
the sum rule (\ref{eq:sumrule}).
Indeed, Eq. (\ref{eq:Ooff}) implies
\begin{equation}
\label{eq:Oab}
O_{\alpha\beta} \sim O(z_1,z_2)\big/R_2(z_1,z_2)
\end{equation}
and hence, from Eq. (\ref{eq:singoldp}),
$O_{\alpha\beta} \sim - N$
if $\lambda_{\alpha}$ and $\lambda_{\beta}$ are neighbouring
eigenvalues in the complex plane, so that (typically) $\omega \sim 1$.

\subsubsection{Distributions of $O_{\alpha\beta}$}
\label{sec:res_dist}
Finally, it is interesting to ask about, not only the average 
behaviour of the overlap matrix, but also its fluctuations. In fact,  
$O_{\alpha\beta}$ is typically large if the matrix $J$ 
has an eigenvalue which is almost degenerate with 
$\lambda_{\alpha}$ or $\lambda_{\beta}$, and as a result, 
the probability distribution of  $O_{\alpha\beta}$ 
has a power-law tail extending to large $|O_{\alpha\beta}|$.
To illustrate this, we consider $N=2$, for
which the probability distribution,
$P(O_{\alpha\alpha})$, of a
 diagonal element of the overlap matrix
is given by Eq. (\ref{eq:dist})
and decays at large $O_{\alpha\alpha}$ according to
$P(O_{\alpha\alpha})\sim O_{\alpha\alpha}^{-3}$.
This implies in particular
that the second and higher moments of $O_{\alpha\alpha}$ diverge.

\begin{figure}
\centerline{\psfig{file=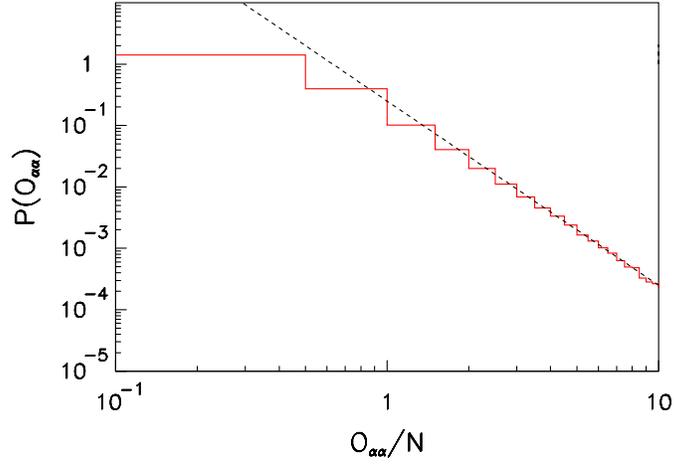,width=10cm}}
\vspace*{1cm}
\caption{\label{fig:hist} A
histogram of   $P(O_{\alpha\alpha})$ of
the diagonal overlaps as a function of $O_{\alpha\alpha}/N$
in Ginibre's ensemble for 
$N=10$ ($-\!\!\!-\!\!\!-\!\!\!-$).
Also shown is the theoretical estimate
for the tail of the distribution  ($- - - $).}
\end{figure}

For $N>2$, the tail of the distribution $P(O_{\alpha\alpha})$
is determined by pairs of eigenvectors with
closest eigenvalues, and we expect that for general $N$, 
the tail of the distribution
function decays algebraically according to 
\begin{equation}
\label{eq:Pasympt}
P(O_{\alpha\alpha}) \sim O_{\alpha\alpha}^{-3}\,.
\end{equation}
In Fig. \ref{fig:hist} we show the distribution $P(O_{\alpha\alpha})$ of
the diagonal overlaps $O_{\alpha\alpha}$
in Ginibre's ensemble for $N=10$. The tail
of the distribution function is well described
by Eq. (\ref{eq:Pasympt}). 

\subsection{Girko's ensemble}
The main result of section \ref{sec:girko} is
Eq. (\ref{eq:Dgirko}), giving  $O(z_1,z_2)$
provided $|z_1-z_2|$ is much greater than the
mean separation in the complex plane
between neighboring eigenvalues.
For $\tau=0$, Girko's ensemble
reduces to Ginibre's ensemble. Correspondingly,
the perturbative result (\ref{eq:Dgirko})
reproduces, for $\tau=0$, $|z_1-z_2| \neq 0$,
$|z_1|,|z_2| < 1$ and large $N$ the expression
(\ref{eq:resO12oldp}), which was obtained
from the exact results of section \ref{sec:ginibre}
in the same limits. The singular contribution
of the diagonal overlap matrix elements to
$D(z_1,z_2)$ is only indirectly available within
perturbation theory. Eq. (\ref{eq:diag}) shows
that the singular behaviour extracted
from the perturbative results is consistent
with the exact expressions
[compare Eq. (\ref{eq:resO11oldp})].
On the other hand, Eq. (\ref{eq:Dgirko})
implies that for $1-\tau \ll 1$,
$O_{\alpha\beta} \sim O(z_1,z_2)/R_2(z_1,z_2) \propto 1-\tau$.
Thus $O_{\alpha\beta}$ vanishes
in the Hermitian limit $\tau \rightarrow 1$, 
as expected. The same is true for the anti-Hermitian
limit, $\tau \rightarrow -1$.

\section{Implications}
\label{sec:imp}
Fluctuations of eigenvectors in
non-Hermitian random matrix ensembles
exhibit a number of striking features
which are likely to be relevant
in physical applications.
As in the immediately preceding sections,
we take the variance in Eqs. (\ref{eq:ginibre})
and (\ref{eq:girko}) to be $\sigma^2=N^{-1}$.

\subsection{Sensitivity to perturbations}
First, as pointed out in the introduction,
systems described  by a non-Hermitian operator
are particularly sensitive to perturbations.
This sensitivity is determined by the diagonal
matrix elements of $O_{\alpha\beta}$.
In order to illustrate this fact, it is convenient
to consider a one-parameter family of matrices 
\begin{equation}
{J} = {J}_1\,\cos\theta + {J}_2\,\sin\theta\,,
\end{equation}
where the parameter  $\theta$ is real and the matrices
${J}_1$ and  ${J}_2$ are drawn independently from the same ensemble.
Then
\begin{equation}
\label{eq:sens}
\left\langle|\partial \lambda_\alpha/\partial \theta|^2\right\rangle
= N^{-1}\,\langle O_{\alpha\alpha}\rangle \,.
\end{equation}
According to Eq. (\ref{eq:resO11oldp}), $\langle O_{\alpha\alpha}\rangle$
is large, being of order $N$. Thus
$\left\langle|\partial \lambda_\alpha/\partial \theta|^2\right\rangle$
is of order unity.  This should be compared
with the Hermitian case \cite{wilk89},
where $\left\langle|\partial \lambda_\alpha/\partial \theta|^2\right\rangle$
is of order $N^{-1}$  in the corresponding parametrization.
Structural stability, on the other hand,  requires
that the level velocities tend to zero
as the boundary of the support of the density of
states is approached: the latter must remain unchanged
as $\theta$ varies, 
since the perturbations merely take $J$ from
one realisation of the ensemble to another.
The expression (\ref{eq:resO11oldp}) for $O(z)$ 
shows that this is indeed the case.

\subsection{Time evolution}
Systems governed
by a non-Hermitian evolution operator may exhibit
transient features in the time-dependence of correlation
functions which are controlled by the type
of correlations between left and right
eigenvectors that we have studied.
Consider for example an evolution equation
of the form
\begin{equation}
\frac{\partial}{\partial t}|u_t\rangle = (J-1)|u_t\rangle
\label{eq:evolution}
\end{equation}
with $J$ drawn from Ginibre's ensemble.
We use $J-1$ rather than $J$ in Eq. (\ref{eq:evolution})
for convenience, to suppress exponential growth.
This corresponds to shifting the support 
of the density of states by unity
along the negative real axis, so that
all (except a vanishing fraction)
of the eigenvalues have negative real parts.
Then    
\begin{equation}
\label{eq:ut}
|u_t\rangle = \sum_{\alpha} |R_\alpha\rangle \,f_t(\lambda_\alpha)\,
  \langle L_\alpha|u_0\rangle\,,
\end{equation}
with $f_t(\lambda)=\exp[(\lambda-1) t]$. Ensemble averaging
with  $\langle u_0 | u_0 \rangle =1$ yields
\begin{equation}
\label{eq:ut2}
\big\langle\langle u_t| u_t\rangle\big\rangle =
\Big\langle\frac{1}{N}\sum_{\alpha\beta} O_{\alpha\beta}\,
f_t(\lambda_\alpha)\,
\bar{f}_t(\lambda_\beta)\Big\rangle\,.
\end{equation}
Thus, properties of the matrix   $O_{\alpha\beta}$
directly influence time evolution.
Eq. (\ref{eq:ut2})  can be obtained as the
double Laplace transform of the density
(\ref{eq:defD}), with respect
to $z_1$ and $z_2$,
\begin{equation}
\label{eq:laplace}
\big\langle\langle u_t| u_t\rangle\big\rangle =
\int\!{\rm d}^2 z_1\,{\rm d}^2 z_2\, {\rm e}^{(z_1 + \overline{z}_2-2) t}\,
     D(z_1,z_2)\,.
\end{equation}
The diagonal and non-diagonal contributions
to $D(z_1,z_2)$ yield large contributions
to Eq. (\ref{eq:laplace}) which almost cancel.
It is thus convenient to evaluate the double
Laplace transform in (\ref{eq:laplace})
by contour integration,
\begin{equation}
\label{eq:contour}
\langle\langle u_t|u_t\rangle\rangle
 = \frac{1}{(2\pi)^2}
   \oint\hspace{-0.85cm}\mbox{\raisebox{-7mm}{$\scriptstyle|z_1|=1$}}
   \!{\rm d}z_1\!
   \oint\hspace{-0.85cm}\mbox{\raisebox{-7mm}{$\scriptstyle|z_2|=1$}}
   \!{\rm d} \overline{z}_2\,\,
   {\rm e}^{(z_1 + \overline{z}_2-2) t} \,
   \big\langle G_{21}(1) G_{12}(2)\big\rangle\,.
\end{equation}
In this case one obtains for large $N$ and for $t\ll\sqrt{N}$ 
\begin{equation}
\label{eq:time1}
\big\langle\langle u_t| u_t\rangle\big\rangle =
{\rm e}^{-2t} \,I_0\big(2 t\big)
\end{equation}
which, for $1 \ll t \ll \sqrt{N}$, simplifies to
\begin{equation}
\label{eq:time1a}
\hspace*{10mm}\sim \frac{1}{\sqrt{4 \pi t}}\,.
\end{equation}
This behaviour should be compared with the much faster decay that would result 
from the same spectrum if the eigenvectors were orthogonal. In the same regime, 
the replacement 
$O_{\alpha\beta} \to \delta_{\alpha\beta}$ transforms Eq.\,(\ref{eq:ut2}) into
\begin{equation}
\big\langle\langle u_t| u_t\rangle\big\rangle =
\Big\langle \frac{1}{N}
\sum_\alpha |f_t(\lambda_\alpha)|^2\Big\rangle
\end{equation}
and thus
\begin{equation}
\big\langle\langle u_t| u_t\rangle\big\rangle =
t^{-1}\,{\rm e}^{-2t}\,I_1\big(2t\big)
\sim \frac{1}{\sqrt{4\pi t^3}}\,.
\end{equation}    
Thus, eigenvector correlations may be as significant as eigenvalue distributions
in determining evolution at intermediate times, 
a fact of established importance in hydrodynamic 
stability theory \cite{tref93,far96}.

\begin{figure}
\centerline{\psfig{file=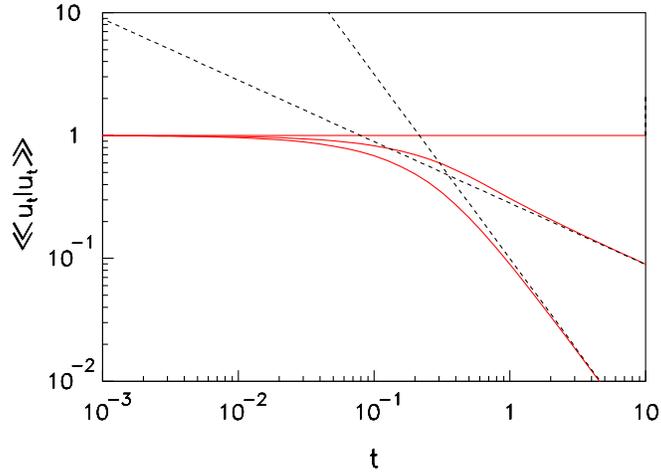,width=10cm,angle=0}}
\vspace*{1cm}
\caption{\label{fig:time} Shows 
$\langle\langle u_t | u_t\rangle\rangle$ 
as a function of $t$ for Girko's ensemble
($-\!\!\!-\!\!\!-\!\!\!-$),
for $\tau=1,0$ and $-1$. Shown
are Eqs. (\protect\ref{eq:time1}) for $\tau=0$
and (\protect\ref{eq:time2}) for $\tau=1$.
For $\tau=-1$, one has just
$\langle\langle u_t | u_t\rangle\rangle=1$.
Also shown are 
the asymptotic expressions valid
for $t \gg 1$ ($- - - $), compare Eq. (\protect\ref{eq:time1a}).
}
\end{figure}

The more general case of Girko's ensemble (\ref{eq:girko})
can also be treated in this way. Mapping the
corresponding contour integrals to the $w$-plane
by means of (\ref{eq:map}), we obtain    
\begin{eqnarray}
\label{eq:contourw}
\langle\langle u_t|u_t\rangle\rangle
  &=& {\rm e}^{-2(1+\tau)t}\,
     \Big\{I_0\big[2(1+\tau)t\big] +\tau\,I_2\big[2(1+\tau)t\big]\Big\}\,.
\end{eqnarray}
Here the support of the spectrum was shifted
by $1+\tau$ along the negative real axis.
The limiting cases of Girko's ensemble, $\tau\rightarrow \pm 1$,
are easily understood: In the anti-Hermitian case,
for $\tau=-1$, one has simply $\langle\langle u_t | u_t \rangle\rangle=1$
because all eigenvalues have vanishing real parts.
In the Gaussian unitary ensemble, for $\tau=1$, on the other hand,
for large $N$, $d(E) = (2\pi)^{-1}\sqrt{4-E^2}$ for $|E| \leq 2$
and thus
\begin{equation}
\label{eq:time2}
\langle\langle u_t | u_t \rangle\rangle
 = \frac{1}{2\pi} \int_{-2}^2\!dE\, \sqrt{4-E^2}\,
{\rm e}^{2(E-2)t}
 = (2t)^{-1}\,{\rm e}^{-4t}\,I_1(4t)\,,
\end{equation}
which corresponds to (\ref{eq:contourw}) for $\tau=1$.
For the three cases,
$\tau=-1,0$ and $1$, $\langle\langle u_t|u_t \rangle\rangle$
is shown in Fig. \ref{fig:time} (full lines) together
with the corresponding
asymptotic expressions (dashed lines) valid for $t \gg 1$.

\subsection{Correlations of eigenvector components}
The space-dependence of correlation functions
of more general ensembles (such as the one discussed in Refs. 
\cite{cha97} and \cite{nel98})
can be modelled by correlation functions of
the components of $\langle L_\alpha|$ and $|R_\beta\rangle$.
Under a change of basis given by a unitary matrix $U$, 
the components of say $|R_\beta\rangle$
transform according to 
$\langle j | R_\beta\rangle \rightarrow \langle j | U|R_\beta\rangle
= \sum_m U_{jm} \langle m|R_\beta\rangle$.
Correspondingly,
$\langle L_\alpha|i\rangle \rightarrow \langle L_\alpha |U^\dagger |i\rangle
= \sum_l \langle L_\alpha | l\rangle \bar {U}_{il} $.
Due to the invariance of the ensemble under
unitary transformations, we can write
\begin{equation}
\langle\langle L_\alpha|i\rangle\langle j|R_\beta\rangle\rangle
= \sum_{ml}
\langle U_{jm} \bar{U}_{il} \rangle_U
\langle L_\alpha|l\rangle\langle m|R_\beta\rangle \,.
\end{equation}
where $\langle \cdots\rangle_U$ denotes an average over 
the unitary matrices $U$.
With $\langle U_{jm} \bar{U}_{il} \rangle_U = N^{-1} \delta_{ij}\delta_{lm}$
this implies immediately
$ \langle\langle L_\alpha|i\rangle\langle j|R_\beta\rangle\rangle
= N^{-1}\delta_{\alpha\beta}\delta_{ij}$.
Consider now averages
involving four 
eigenvector components. The only non-vanishing
(and non-trivial)
averages which are invariant under the 
scale transformation (\ref{eq:scaletr}) are
\begin{equation}
\label{eq:3}
\langle
\langle i|R_\alpha\rangle\langle L_\alpha|j\rangle
\langle j|L_\beta \rangle\langle R_\beta | i\rangle
\rangle 
= \frac{1}{N^2-1} \big(\delta_{ij} + \langle O_{\alpha\beta}\rangle\big)
-\frac{1}{N}\frac{1}{N^2-1}\big(1+\delta_{ij}\langle O_{\alpha\beta}\rangle\big)
\end{equation}
and
\begin{equation}
\label{eq:4}
\langle
\langle i|R_\alpha\rangle\langle L_\alpha|i\rangle
\langle j|L_\beta \rangle\langle R_\beta | j\rangle
\rangle 
= \frac{1}{N^2-1} \big(1+\delta_{ij}\langle O_{\alpha\beta}\rangle\big)
-\frac{1}{N}\frac{1}{N^2-1}\big(\delta_{ij}
+\langle O_{\alpha\beta}\rangle\big)\,.
\end{equation}
Summing Eqs. (\ref{eq:3}) and (\ref{eq:4})
over $i$ and $j$ one obtains $\langle O_{\alpha\beta}\rangle$
and unity, respectively, as expected.
It should be noted that the dependence on $i,j$ and
$\alpha,\beta$ does not necessarily factorize.
This is likely to be of importance in 
problems with spatial structure. The above
considerations show that 
interesting space dependence of correlation functions may
arise from non-Hermiticity.

\section{Conclusions}
\label{sec:conc}
In this paper
we have analyzed correlations of eigenvectors
in non-Hermitian random matrix ensembles. 
Such correlations are of interest
partly because they determine
some aspects of the behavior of 
systems represented by non-Hermitian
operators: for example, such systems are particularly sensitive to external
perturbations and correlation functions may
exhibit transient features in their time dependences.
As emphasized 
in the introduction,
there are numerous instances in which random
non-Hermitian operators appear in the description of physical
problems, and we hope that the results and methods
summarized here will 
be of interest in a number
of contexts.
In particular, we have
obtained the following results.
We have characterized exactly the
eigenvector correlations in
Ginibre's ensemble of non-Hermitian random
matrices.
We have shown that the sensitivity
of the eigenvalues with respect
to external perturbations
is larger by a factor of $N$ (where $N$ is the rank of the matrix)
than the equivalent for Dyson's ensembles of
Hermitian matrices.
Moreover, we have shown that
eigenvectors associated with two different
eigenvalues exhibit strong correlations
which decrease algebraically with increasing
separation between the eigenvalues in
the complex plane. We have also
shown that the probability distribution function
of eigenvector overlaps has
algebraic tails. This implies that
fluctuations are large, in the sense that higher moments
of the eigenvector overlaps diverge.
In addition to exact calculations specific to 
Ginibre's ensemble, an alternative, perturbative approach has been
developed and used
to derive corresponding results, in
an approximate way, in Girko's more
general ensemble of non-Hermitian random
matrices. In the appropriate cases and in
the limit of large $N$, the exact results are reproduced.

\acknowledgements{BM gratefully acknowledges support of the SFB393. JTC is supported in part by EPSRC Grant No. GR/MO4426.}

\newpage

\end{document}